 \documentstyle[preprint,pre,aps,eqsecnum,epsf,amssymb]{revtex}
 \tighten

\begin{document}
\draft
\title{Scaling Limit of the Ising Model in a Field}
\author{Uwe Grimm\cite{email:ug,oldaddress:ug}}
\address{Institut f\"{u}r Physik,
Technische Universit\"{a}t Chemnitz,\\
09107 Chemnitz, Germany}
\author{Bernard Nienhuis\cite{email:bn}}
\address{Instituut voor Theoretische Fysica,
Universiteit van Amsterdam,\\
Valckenierstraat 65, 
1018 XE Amsterdam, The Netherlands}
\date{\today}
\maketitle
\begin{abstract}
The dilute $\mbox{A}_3$ model is a solvable IRF (interaction round a face)
model with three local states and adjacency conditions encoded by
the Dynkin diagram of the Lie algebra $\mbox{A}_3$. It can be regarded as a 
solvable version of an Ising model at the critical temperature in a 
magnetic field. One therefore expects the scaling limit to be governed
by Zamolodchikov's integrable perturbation of the \mbox{$c=1/2$} conformal
field theory. Indeed, a recent thermodynamic Bethe Ansatz approach succeeded
to unveil the corresponding $\mbox{E}_{8}$ structure under certain 
assumptions on the nature of the 
Bethe Ansatz solutions. In order to check these conjectures, we perform
a detailed numerical investigation of the solutions of the Bethe Ansatz 
equations for the critical and off-critical model. 
Scaling functions for the ground-state corrections and for the lowest 
spectral gaps are obtained, which give very precise numerical results
for the lowest mass ratios in the massive scaling limit. While these
agree perfectly with the $\mbox{E}_{8}$ mass ratios, we observe one
state which seems to violate the assumptions underlying the 
thermodynamic Bethe Ansatz calculation. We also analyze the
critical spectrum of the dilute $\mbox{A}_3$ model, 
which exhibits massive excitations on top of the massless 
states of the Ising conformal field theory.
\end{abstract}




\narrowtext
\section{Introduction}
\label{s:int}

The Ising model \cite{Ising} is, without doubt, one of the
most frequently studied and best understood lattice models of
classical statistical mechanics. Although Onsager's 
solution~\cite{Onsager} of the two-dimensional Ising model without 
external field dates back half a century already, no analytic solution
of the 2D Ising model in a magnetic field has been found.

However, the situation is somewhat different if considered 
from the viewpoint of field theory. The critical 2D Ising
model corresponds to the $c=1/2$ conformal field 
theory (CFT)~\cite{BPZ,Cardy1,Ginsparg,Cardy2} of a 
massless Majorana fermion.
It was Zamolodchikov~\cite{Zamo87,Zamo89a,Zamo89b} who noticed
that a (symmetry-breaking)
perturbation of this CFT with the
relevant spin density operator (which has conformal dimensions 
$(\Delta,\overline{\Delta})=(1/16,1/16)$) preserves infinitely
many conservation laws and therefore leads to an integrable
quantum field theory. The corresponding minimal theory
contains eight massive particles with factorized (purely elastic) 
scattering~\cite{ZamoZamo,KoebSwie,OgieReshWieg,FatZamo,KlassMelz90,Mussardo}; 
the particle masses and the $S$-matrix elements are related 
to the exceptional Lie algebra $\mbox{E}_8$. 

This integrable field theory describes an appropriate
scaling limit of the 2D Ising model in a magnetic field.
Numerically, the predictions for the lowest mass ratios have 
been verified by several 
authors~\cite{Henkel87,SagZamo,HenkSal,LauwRitt,Henkel91}.
As these results rely on relatively small size transfer matrix 
calculations or on Monte Carlo simulations, they provide
rather crude checks for the lowest mass ratios only. Furthermore, 
the larger masses cannot be obtained directly as these lie above the 
two-particle threshold of the lightest particles and hence
are buried in the continuum of scattering states. 
However, one can extract indirect 
information about the higher masses and the scattering amplitudes
from the finite-size behavior of the lower 
masses~\cite{KlassMelz91,Henkel91}.
Another, rather different, approach employs a truncated
conformal Hilbert space~\cite{YurZam90,YurZam91}, which however
does not make the connection to the lattice model. 

Similar theoretical and numerical investigations have also been 
performed for a variety of lattice models such as 
the quantum Ising chain~\cite{Henkel87},
the Lee-Yang model~\cite{YurZam90}, 
the three-state Potts model~\cite{Zamo88}, 
the tricritical Ising respectively 
Blume-Capel model~\cite{Henkel90,Gehlen,LMC91},
the Ashkin-Teller model~\cite{HenLud}, 
the $Z(N)$ models~\cite{Alcaraz}, and the
integrable RSOS models~\cite{BazhResh}. Recently,
this has been extended to the study of form factors and correlation 
functions~\cite{DelMus,GuiMag96a,AceMusVal,DelSim,GuiMag96b}
and non-integrable perturbations~\cite{DelMusSim}.

The discovery of a new class of lattice 
models~\cite{Roche92,WNS92,WN93}
which are solvable in the presence of a symmetry-breaking 
field~\cite{WNS92} has changed the situation considerably. Although we
still cannot solve the 2D Ising model in a magnetic field,
we now know a solvable model (the so-called dilute $\mbox{A}_3$ model)
which belongs to the same universality class: it has a critical point 
of Ising type~\cite{WNS92,WNS93}, and it is solvable
in a symmetry-breaking field that corresponds to the
spin density perturbation of the CFT~\cite{WPNS94}. 
Therefore, one expects this model
to show the same properties in the scaling limit as the
non-integrable Ising model in a magnetic field. Note, however,
that the two models are rather different lattice models: the
dilute $\mbox{A}_3$ model contains three local states, whereas
the Ising spins take two values only. 

The dilute $\mbox{A}_L$ models are solvable by Bethe 
Ansatz (BA) \cite{BNW94,ZPG95}. By a thermodynamic Bethe Ansatz (TBA)
approach~\cite{BNW94}, the mass ratios and $S$-matrix of the
$\mbox{E}_8$ field theory have been obtained. However, this
approach relies on assumptions on the nature of the BA
solutions, which could not be substantiated by numerical
solution of the BA equations (BAE) at criticality.
Connections between the $\mbox{A}_3$ model
and the exceptional Lie algebra $\mbox{E}_8$ are
also revealed by the existence of corresponding
Rogers-Ramanujan identities~\cite{WP94}.

For our numerical investigation of the off-critical 
BAE, our main motivation was to check 
the assumptions on the conjectured string structure of the
BA solutions. This could be achieved for the states 
with lowest masses, yielding at the same time very accurate
numerical results not alone for the mass ratios, but also
for the complete scaling functions. However, we also found
{\em one discrepancy\/} with the predictions of Ref.~\cite{BNW94}.
Within the range of scaling parameter values considered, 
the characteristic string type for a particle 
of mass $m_4$ does not match the proposed structure. 
 
As a by-product of our investigation, we found that the 
critical spectrum also 
contains {\em massive\/} excitations besides the well-known 
conformal spectrum. This is one property that distinguishes
the dilute $\mbox{A}_3$ model from the Ising model which does
not have such excitations. However, this does not contradict
universality as that only concerns the universal
properties of the systems in the vicinity of the critical point, 
which are the same for both models. The BA solutions 
corresponding to the massive excitations show the 
same type of strings as those conjectured for the massive 
scaling limit. Already for relatively small systems, these
solutions lie very close to singularities of the BAE, which 
makes the numerical treatment difficult. Estimates of the 
corresponding mass ratios, which show no resemblance of 
the $\mbox{E}_8$ structure, are presented.

This paper is organized as follows. In Section~\ref{s:A3}, 
the dilute $\mbox{A}_3$ models is introduced and the BAE
are presented. The critical spectrum predicted by
CFT and the behavior in the massive scaling
limit are described. Furthermore, the assumptions on the form
of the BA solutions formulated in Ref.~\cite{BNW94} are
summarized.

Section~\ref{s:nr} contains the results of our numerical investigations.
After a brief description of our numerical approach, we commence
with the critical case, where not only the scaling dimensions
predicted by CFT are considered, but also
additional massive excitations in the spectrum are identified.
Then, we show how the BA solutions behave as one goes 
away from criticality. Finally, we present our numerical scaling 
functions for the lowest gaps and the numerical results for the 
mass ratios in the massive scaling limit.

\section{The Dilute $\mbox{A}_3$ Model}
\label{s:A3}

Because an immediate specialization to the $\mbox{A}_3$ case does not 
considerably simplify the equations, we keep the first part of this 
section on a more general level and consider the dilute $\mbox{A}_L$ 
models. 
This also serves us to avoid an unnecessary repetition in a 
forthcoming publication~\cite{GN}. 

\begin{figure}
\caption{Effective adjacency diagram of the dilute $\mbox{A}_3$ model.}
\label{f:ead}\nopagebreak
\centerline{\epsfxsize=0.7\columnwidth \epsfbox{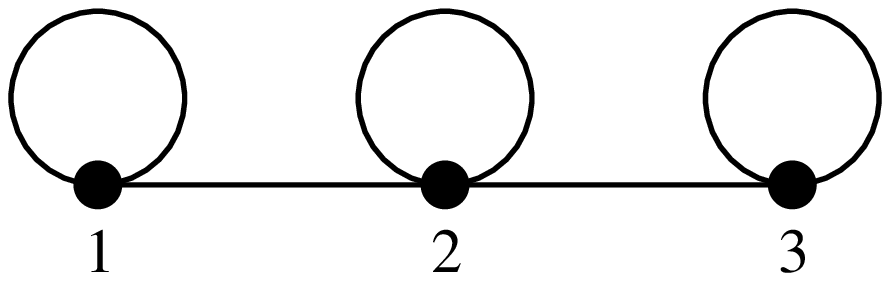}}
\end{figure}

The dilute $\mbox{A}_L$ models are IRF (interaction round a face)
models~\cite{Baxter} of RSOS (restricted solid-on-solid) 
type~\cite{ABF} on the square lattice with adjacency conditions
encoded by the Dynkin diagram of the Lie algebra $\mbox{A}_L$.
In contrast to the usual (non-dilute) RSOS models built on this
adjacency graph (these are the ABF models \cite{ABF}), the effective
adjacency graph of the dilute $\mbox{A}_L$ models contains
loops which connect each node to itself, see Fig.~\ref{f:ead}.
In other words, one considers configurations of $L$ local states 
or ``heights'' (labeled $1,2,\ldots,L$) on the vertices of the
square lattice subject to the additional requirement that states on
neighboring (adjacent) lattice sites may differ {\em at most\/} by one. 
One then defines a statistical model by assigning Boltzmann weights to 
the elementary plaquettes (faces) of the lattice. 

The face weights of the dilute $\mbox{A}_L$ models are \cite{WNS92} 
\widetext
\ifpreprintsty
\begin{eqnarray}
W\!\left(\matrix{a&a\cr a&a\cr}\biggm| u\right) &\; =\; &
\frac{\vartheta_{1}(6\lambda-u)\,\vartheta_{1}(3\lambda+u)}
{\vartheta_{1}(6\lambda)\,\vartheta_{1}(3\lambda)} \nonumber\\* 
& & -\:\left(\frac{S_{a+1}}{S_a}\,
\frac{\vartheta_{4}(2a\lambda-5\lambda)}{\vartheta_{4}(2a\lambda+\lambda)}
 +\frac{S_{a-1}}{S_a}\,
\frac{\vartheta_{4}(2a\lambda+5\lambda)}
{\vartheta_{4}(2a\lambda-\lambda)}\right)
\frac{\vartheta_{1}(u)\,\vartheta_{1}(3\lambda-u)}
{\vartheta_{1}(6\lambda)\,\vartheta_{1}(3\lambda)} \nonumber\\ 
W\!\left(\matrix{a\!\pm\! 1&a\cr a&a\cr}\biggm| u\right) & = &
W\!\left(\matrix{a&a\cr a&a\!\pm\! 1\cr}\biggm| u\right) \;\; = \;\;
\frac{\vartheta_{1}(3\lambda-u)\,\vartheta_{4}(\pm 2a\lambda+\lambda-u)}
{\vartheta_{1}(3\lambda)\,\vartheta_{4}(\pm 2a\lambda+\lambda)}\nonumber\\
W\!\left(\matrix{a&a\cr a\!\pm\! 1&a\cr}\biggm| u\right) & = &
W\!\left(\matrix{a&a\!\pm\! 1\cr a&a\cr}\biggm| u\right) \;\; = \;\; 
\left(\frac{S_{a\pm 1}}{S_a}\right)^{1/2}\,
\frac{\vartheta_{1}(u)\,\vartheta_{4}(\pm 2a\lambda-2\lambda+u)}
{\vartheta_{1}(3\lambda)\,\vartheta_{4}(\pm 2a\lambda+\lambda)} \nonumber\\ 
W\!\left(\matrix{a&a\!\pm\! 1\cr a&a\!\pm\! 1\cr}\biggm| u\right) & = &
W\!\left(\matrix{a\!\pm\! 1&a\!\pm\! 1\cr a&a\cr}\biggm| u\right)\nonumber\\
&  = &\left(\frac{\vartheta_{4}(\pm 2a\lambda+3\lambda)\,
\vartheta_{4}(\pm 2a\lambda-\lambda)}
{\vartheta_{4}^{2}(\pm 2a\lambda+\lambda)}\right)^{1/2}\,
\frac{\vartheta_{1}(u)\,\vartheta_{1}(3\lambda-u)}
{\vartheta_{1}(2\lambda)\,\vartheta_{1}(3\lambda)} \nonumber\\
W\!\left(\matrix{a\!\pm\! 1&a\cr a&a\!\mp\! 1\cr}\biggm| u\right) & = &
\frac{\vartheta_{1}(2\lambda-u)\,\vartheta_{1}(3\lambda-u)}
{\vartheta_{1}(2\lambda)\,\vartheta_{1}(3\lambda)} \nonumber\\ 
W\!\left(\matrix{a&a\!\mp\! 1\cr a\!\pm\! 1&a\cr}\biggm| u\right) & = &
-\left(\frac{S_{a-1}S_{a+1}}{S^2_a}\right)^{1/2}\,
\frac{\vartheta_{1}(u)\,\vartheta_{1}(\lambda-u)}
{\vartheta_{1}(2\lambda)\,\vartheta_{1}(3\lambda)} \nonumber\\
W\!\left(\matrix{a&a\!\pm\! 1\cr a\!\pm\! 1&a\cr}\biggm| u\right) & = &
\frac{\vartheta_{1}(3\lambda-u)\,\vartheta_{1}(\pm 4a\lambda+2\lambda+u)}
{\vartheta_{1}(3\lambda)\,\vartheta_{1}(\pm 4a\lambda+2\lambda)}
+\frac{S_{a\pm 1}}{S_a}\,
\frac{\vartheta_{1}(u)\,\vartheta_{1}(\pm 4a\lambda-\lambda+u)}
{\vartheta_{1}(3\lambda)\,\vartheta_{1}(\pm 4a\lambda+2\lambda)} \nonumber\\*
& = & 
\frac{\vartheta_{1}(3\lambda+u)\,\vartheta_{1}(\pm 4a\lambda-4\lambda+u)}
{\vartheta_{1}(3\lambda)\,\vartheta_{1}(\pm 4a\lambda-4\lambda)} \nonumber\\*
& & +\:\left(\frac{S_{a\mp 1}}{S_a}\,
\frac{\vartheta_{1}(4\lambda)}{\vartheta_{1}(2\lambda)}
-\frac{\vartheta_{4}(\pm 2a\lambda-5\lambda)}
{\vartheta_{4}(\pm 2a\lambda+\lambda)} \right)
\frac{\vartheta_{1}(u)\,\vartheta_{1}(\pm 4a\lambda-\lambda+u)}
{\vartheta_{1}(3\lambda)\,\vartheta_{1}(\pm 4a\lambda-4\lambda)}\nonumber\\
\label{e:fw}
\end{eqnarray}
\else
\begin{eqnarray}
W\!\left(\matrix{a&a\cr a&a\cr}\biggm| u\right) &\; =\; &
\frac{\vartheta_{1}(6\lambda-u)\,\vartheta_{1}(3\lambda+u)}
{\vartheta_{1}(6\lambda)\,\vartheta_{1}(3\lambda)} \nonumber\\* 
& & -\:\left(\frac{S_{a+1}}{S_a}\,
\frac{\vartheta_{4}(2a\lambda-5\lambda)}{\vartheta_{4}(2a\lambda+\lambda)}
 +\frac{S_{a-1}}{S_a}\,
\frac{\vartheta_{4}(2a\lambda+5\lambda)}
{\vartheta_{4}(2a\lambda-\lambda)}\right)
\frac{\vartheta_{1}(u)\,\vartheta_{1}(3\lambda-u)}
{\vartheta_{1}(6\lambda)\,\vartheta_{1}(3\lambda)} \nonumber\\ 
W\!\left(\matrix{a\!\pm\! 1&a\cr a&a\cr}\biggm| u\right) & = &
W\!\left(\matrix{a&a\cr a&a\!\pm\! 1\cr}\biggm| u\right) \;\; = \;\;
\frac{\vartheta_{1}(3\lambda-u)\,\vartheta_{4}(\pm 2a\lambda+\lambda-u)}
{\vartheta_{1}(3\lambda)\,\vartheta_{4}(\pm 2a\lambda+\lambda)}\nonumber\\
W\!\left(\matrix{a&a\cr a\!\pm\! 1&a\cr}\biggm| u\right) & = &
W\!\left(\matrix{a&a\!\pm\! 1\cr a&a\cr}\biggm| u\right) \;\; = \;\; 
\left(\frac{S_{a\pm 1}}{S_a}\right)^{1/2}\,
\frac{\vartheta_{1}(u)\,\vartheta_{4}(\pm 2a\lambda-2\lambda+u)}
{\vartheta_{1}(3\lambda)\,\vartheta_{4}(\pm 2a\lambda+\lambda)} \nonumber\\ 
W\!\left(\matrix{a&a\!\pm\! 1\cr a&a\!\pm\! 1\cr}\biggm| u\right) & = &
W\!\left(\matrix{a\!\pm\! 1&a\!\pm\! 1\cr a&a\cr}\biggm| u\right) 
\;\; = \;\;\left(\frac{\vartheta_{4}(\pm 2a\lambda+3\lambda)\,
\vartheta_{4}(\pm 2a\lambda-\lambda)}
{\vartheta_{4}^{2}(\pm 2a\lambda+\lambda)}\right)^{1/2}\,
\frac{\vartheta_{1}(u)\,\vartheta_{1}(3\lambda-u)}
{\vartheta_{1}(2\lambda)\,\vartheta_{1}(3\lambda)} \nonumber\\
W\!\left(\matrix{a\!\pm\! 1&a\cr a&a\!\mp\! 1\cr}\biggm| u\right) & = &
\frac{\vartheta_{1}(2\lambda-u)\,\vartheta_{1}(3\lambda-u)}
{\vartheta_{1}(2\lambda)\,\vartheta_{1}(3\lambda)} \nonumber\\ 
W\!\left(\matrix{a&a\!\mp\! 1\cr a\!\pm\! 1&a\cr}\biggm| u\right) & = &
-\left(\frac{S_{a-1}S_{a+1}}{S^2_a}\right)^{1/2}\,
\frac{\vartheta_{1}(u)\,\vartheta_{1}(\lambda-u)}
{\vartheta_{1}(2\lambda)\,\vartheta_{1}(3\lambda)} \nonumber\\
W\!\left(\matrix{a&a\!\pm\! 1\cr a\!\pm\! 1&a\cr}\biggm| u\right) & = &
\frac{\vartheta_{1}(3\lambda-u)\,\vartheta_{1}(\pm 4a\lambda+2\lambda+u)}
{\vartheta_{1}(3\lambda)\,\vartheta_{1}(\pm 4a\lambda+2\lambda)}
+\frac{S_{a\pm 1}}{S_a}\,
\frac{\vartheta_{1}(u)\,\vartheta_{1}(\pm 4a\lambda-\lambda+u)}
{\vartheta_{1}(3\lambda)\,\vartheta_{1}(\pm 4a\lambda+2\lambda)} \nonumber\\*
& = & 
\frac{\vartheta_{1}(3\lambda+u)\,\vartheta_{1}(\pm 4a\lambda-4\lambda+u)}
{\vartheta_{1}(3\lambda)\,\vartheta_{1}(\pm 4a\lambda-4\lambda)} \nonumber\\*
& & +\:\left(\frac{S_{a\mp 1}}{S_a}\,
\frac{\vartheta_{1}(4\lambda)}{\vartheta_{1}(2\lambda)}
-\frac{\vartheta_{4}(\pm 2a\lambda-5\lambda)}
{\vartheta_{4}(\pm 2a\lambda+\lambda)} \right)
\frac{\vartheta_{1}(u)\,\vartheta_{1}(\pm 4a\lambda-\lambda+u)}
{\vartheta_{1}(3\lambda)\,\vartheta_{1}(\pm 4a\lambda-4\lambda)} 
\label{e:fw}
\end{eqnarray}
\fi 
\narrowtext
Here, $a=1,2,\ldots L$ labels the heights, and
the possible values of the variable $\lambda$ are 
determined by $L$ \cite{WNS92}. 
The crossing factors $S_a$ are given by
\begin{equation}
S_a \;\; = \;\; (-1)^{\displaystyle a} \; 
\frac{\vartheta_{1}(4a\lambda)}{\vartheta_{4}(2a\lambda)} 
\end{equation}
and $\vartheta_{1}(u)$, $\vartheta_{4}(u)$ are standard 
theta functions of nome $q$ with $|q|<1$ \cite{GradRyz}. 
These face weights satisfy the Yang-Baxter equation and
therefore lead to an integrable lattice model. The corresponding
row transfer matrices~\cite{Baxter} for fixed value of $q$ 
form a one-parameter commuting family in the spectral parameter $u$.

In general, the dilute $\mbox{A}_L$ models have four different branches 
for any value of $L$ \cite{WNS92}, distinguished by two possible values
of $\lambda$ with two regimes each for the spectral parameter $u$. 
Here, we are only interested in the dilute $\mbox{A}_3$ model ($L=3$)
in a particular branch \cite{WNS92,WNS93,BNW94} where the model behaves
like the Ising model in a magnetic field.  For this case, $\lambda=5\pi/16$ 
and the spectral parameter $u$ lies in the interval $0<u<3\lambda$.
It is the parameter $q$ that acts like the magnetic field, and the 
Ising critical point corresponds to $q=0$. 

\subsection{Bethe Ansatz Equations}
\label{ss:bae}

The eigenvalues $\Lambda(u)$ of the row transfer matrix for a system of 
size $N$ with periodic boundary conditions have the 
following form~\cite{BNW94,ZPG95} 
\widetext
\begin{eqnarray}
\Lambda(u) &\; =\; &
\omega\;{\left(-\,
\frac{\vartheta_{1}(u-2\lambda)\,\vartheta_{1}(u-3\lambda)}
{\vartheta_{1}(2\lambda)\,\vartheta_{1}(3\lambda)}\right)}^N
\;\prod_{j=1}^N 
\frac{\vartheta_{1}(u-u_j+\lambda)}{\vartheta_{1}(u-u_j-\lambda)}
 \nonumber\\* & & +\;{\left(-\,
\frac{\vartheta_{1}(u)\,\vartheta_{1}(u-3\lambda)}
{\vartheta_{1}(2\lambda)\,\vartheta_{1}(3\lambda)}\right)}^N
\;\prod_{j=1}^N
\frac{\vartheta_{1}(u-u_j)\,\vartheta_{1}(u-u_j-3\lambda)}
{\vartheta_{1}(u-u_j-\lambda)\,\vartheta_{1}(u-u_j-2\lambda)}
 \nonumber\\* &  & +\;\omega^{-1}\; {\left(-\,
\frac{\vartheta_{1}(u)\,\vartheta_{1}(u-\lambda)}
{\vartheta_{1}(2\lambda)\,\vartheta_{1}(3\lambda)}\right)}^N \;
\prod_{j=1}^N 
\frac{\vartheta_{1}(u-u_j-4\lambda)}{\vartheta_{1}(u-u_j-2\lambda)}
\label{e:eval}
\end{eqnarray}
where $\omega=\exp(i\pi/(L+1))$, and where
the $u_j$, $j=1,\ldots,N$, form a solution of the set of $N$ coupled BAE 
\begin{equation}
\omega^{\ell}\;{\left(
\frac{\vartheta_{1}(u_j-\lambda)}
{\vartheta_{1}(u_j+\lambda)}\right)}^{N} 
\;\; = \;\; -\,\prod_{k=1}^{N}
\frac{\vartheta_{1}(u_j-u_k-2\lambda)\,\vartheta_{1}(u_j-u_k+\lambda)}
{\vartheta_{1}(u_j-u_k+2\lambda)\,\vartheta_{1}(u_j-u_k-\lambda)}
\label{e:BAE}
\end{equation}
\narrowtext
Here, $\ell=1,\ldots,L$ labels a sector related to the braid limit 
eigenvalues of the row transfer matrix at criticality ($q=0$).
Whereas it is believed to be true that all eigenvalues of the 
transfer matrix are of the form (\ref{e:eval}), the converse is 
certainly wrong (at least off-criticality) --- the BAE (\ref{e:BAE}) 
in general allow for many additional solutions which do not
correspond to proper eigenvalues of the transfer matrix.
For any numerically found solution of Eq.~(\ref{e:BAE})
one thus has to check whether it gives a proper eigenvalue of 
the transfer matrix (see Section~\ref{s:nr} below). 

The theta function $\vartheta_{1}(u)$ shows 
the (quasi-)periodicity properties \cite{GradRyz} 
\begin{mathletters}
\begin{eqnarray}
\vartheta_{1}(u+\pi) &\; =\; & -\,\vartheta_{1}(u) 
\;\; = \;\; \vartheta_{1}(-u) \label{e:per} \\
\vartheta_{1}(u+\pi\tau) & = & 
-\,\frac{1}{q}\: e^{-2iu}\: \vartheta_{1}(u) \label{e:quasi}
\end{eqnarray}
\end{mathletters}%
where $q=\exp(i\pi\tau)$ ($0<q<1$) with $\tau\in i\mathbb{R}$. 
Consider a solution 
$\{u_1,\ldots,u_N\}$ of the BAE (\ref{e:BAE}) in sector $\ell$.
Clearly, nothing is changed if a multiple of $\pi$ is added to 
any of the roots $u_j$.

More interesting is the result of adding
$n\pi\tau$ ($n\in\mathbb{Z}$) to a root $u_j$. 
In this case, one obtains 
a solution of (\ref{e:BAE}) with $\omega^{\ell}$ replaced by 
$\exp(4ni\lambda)\omega^{\ell}$.
Moreover, from Eq.~(\ref{e:eval}) it is evident 
that these two solutions correspond to the
same eigenvalue. This means that the ``sectors'' 
$\ell=1,\ldots,L$ lose their
significance in the off-critical case 
($q\neq 0$) since we can always
adjust $\ell$ by adding or subtracting 
suitable multiples of $\pi\tau$
to some of the roots. Of course, this 
also yields many possibilities
in which the phase factors cancel, and hence 
Eq.~(\ref{e:BAE}) is recovered
without any alteration, which means that 
a single solution of the
BAE can be presented in many ways differing by
addition and subtraction of suitable 
multiples of $\pi\tau$ to some of the roots.

\begin{figure}
\caption{Sketch of a typical arrangement of the Bethe Ansatz 
         roots $v_{j}$ in the complex plane.}
\label{f:string}\nopagebreak
\centerline{\epsfxsize=\columnwidth \epsfbox{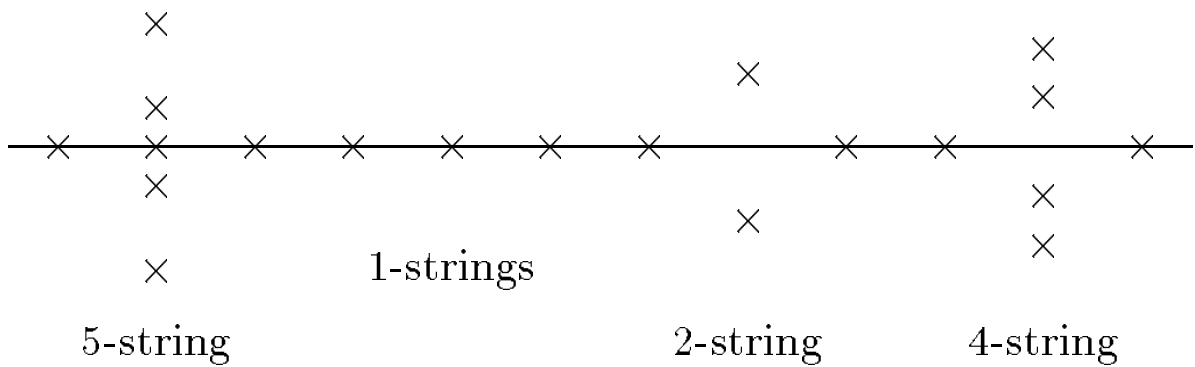}}
\end{figure}

For the case of interest, it turns out that the largest eigenvalue
$\Lambda_{0}^{}(u)$ of the transfer matrix is given by a purely 
imaginary solution of the BAE (\ref{e:BAE}) in the sector $\ell=1$. 
Therefore, in place of the
roots $u_j$, we prefer to use $v_j=iu_j$ in what follows; thus the
largest eigenvalue corresponds to a set of real BA roots $v_j$.
General solutions to the BAE will however involve complex roots,
which in the large $N$ limit typically arrange into so-called 
{\em strings\/} (subsets of roots with approximately the same 
real part, in terms of the $v_j$). 
This is shown schematically in Fig.~\ref{f:string}, where
the horizontal line represents the real axis. Note that the
``string content'' of a particular solution  might well depend on
the variable $q$, an example of this behavior will be given 
in Section~\ref{ss:bar} below.

\subsection{Conformal Spectrum at Criticality}
\label{ss:cs}

The dilute $\mbox{A}_3$ model has a critical point of 
Ising type at $q=0$. Thus its critical limit (given
by $q=0$, $N\rightarrow\infty$) is described 
by the $c=1/2$ CFT with scaling 
dimensions $\Delta\in\{0,1/16,1/2\}$. This has drastic 
consequences for the spectrum of the transfer matrix in 
the critical limit. Consider the scaled spectral gaps
\begin{equation}
x_{j}^{} \; = \; \frac{N}{2\pi}\,
\log(\Lambda_{0}^{(0)}/\Lambda_{j}^{(0)})\; ,
\label{e:xj}
\end{equation}
where $\Lambda_{j}^{(0)}$ are the eigenvalues of the 
transfer matrix with periodic boundary conditions 
at the isotropic point $u=3\lambda/2$, 
$\Lambda_{0}^{(0)}$ being the largest eigenvalue.
In the critical limit, the spectrum of scaling dimensions $x_j$ 
consists of ``conformal towers'' of states labeled by pairs
$(\Delta+r,\overline{\Delta}+\overline{r})$ with scaled energy
$x_j=\Delta+r+\overline{\Delta}+\overline{r}$, 
conformal spin $s_j=\Delta-\overline{\Delta}$
and momentum $p_j=r-\overline{r}$, 
where $r,\overline{r}\in{\mathbb{N}}_{0}$.
They form representations of two commuting Virasoro algebras with
central charge $c=1/2$. The degeneracies of the descendant states
can be read off from the character functions
of their irreducible representations with highest weight $\Delta$.
These can be written in the form \cite{Baake88}
\begin{mathletters} 
\begin{eqnarray}
\chi_{0}^{}(z) &\; =\; & \sum_{n\in\mathbb{Z}}
     z^{4n^2+n} \:\Pi_{V}^{}(z^2) \nonumber \\
 & = &  1 + z^2 + z^3 + 2 z^4 + 2 z^5 + \ldots \label{e:chi0} \\
\chi_{\frac{1}{2}}^{}(z) & = & \sum_{n\in\mathbb{Z}}
     z^{4n^2+3n+\frac{1}{2}} \:\Pi_{V}^{}(z^2) \nonumber \\
 & = & z^{\frac{1}{2}}\, (1 + z + z^2 + z^3 + 2 z^4 + \ldots )
 \label{e:chi12} \\
\chi_{\frac{1}{16}}^{}(z) & = & z^{\frac{1}{16}}
     \prod_{m=1}^{\infty}(1+z^{m}) \nonumber \\ 
& = & z^{\frac{1}{16}}\, (1 + z + z^2 + 2 z^3 + 2 z^4 + \ldots )
\label{e:chi116} 
\end{eqnarray}
\end{mathletters}%
where $\Pi_{V}^{}(z)$ is the generating function 
of the number of partitions
\begin{equation}
\Pi_{V}^{}(z) = \prod_{m=1}^{\infty} \frac{1}{1-z^m} \; .
\end{equation}
Moreover, the central charge $c=1/2$ manifests itself in the
finite-size corrections of the largest eigenvalue~\cite{BCN}
\begin{equation}
-\log(\Lambda_{0}^{(0)}) \; = \;
N f_{0}\, + \,\frac{\pi c}{6 N}\, +\, o(N^{-1}) 
\label{e:cc}
\end{equation}
where $f_{0}$ denotes the bulk free energy for $u=3\lambda/2$.

\subsection{Scaling Limit}
\label{ss:sl}

Taking into consideration the parameter $q$, 
one can approach the critical point by simultaneously
performing the two limits $q\rightarrow 0$ and 
$N\rightarrow\infty$ keeping the scaling variable 
\begin{equation}
\mu\;=\; q\: N^{15/8}
\label{e:mu}
\end{equation}
constant. Here, $\mu=0$ corresponds to the critical limit discussed 
above, $\mu\rightarrow\infty$ to the massive limit where 
Zamolodchikov's $\mbox{E}_8$ field theory results apply. 
In the scaling limit, the appropriately scaled spectral gaps 
\begin{equation}
F_{j} \; = \; q^{-8/15}\: \log(\Lambda_{0}^{}/\Lambda_{j}^{})
\label{e:fj}
\end{equation}
become functions of the scaling variable $\mu$ alone.
For the largest eigenvalue, we can also define a scaling function 
\begin{equation}
F_{0} \; = \; q^{-8/15}\: \left[\log(\Lambda_{0}^{})+N\cdot f_0\right]
\label{e:f0}
\end{equation}
for the finite-size corrections of the largest eigenvalue,
where $f_0=\lim_{N\rightarrow\infty}[-\log(\Lambda_{0}^{})/N]$ 
is the bulk free energy again.
Note that the face weights (\ref{e:fw}) for nome $q$ and $-q$ are
related by symmetry, thus it suffices to consider positive values
of the nome.

\begin{figure}
\caption{The $\mbox{E}_8$ Dynkin diagram}
\label{f:dd}\nopagebreak
\centerline{\epsfxsize=\columnwidth \epsfbox{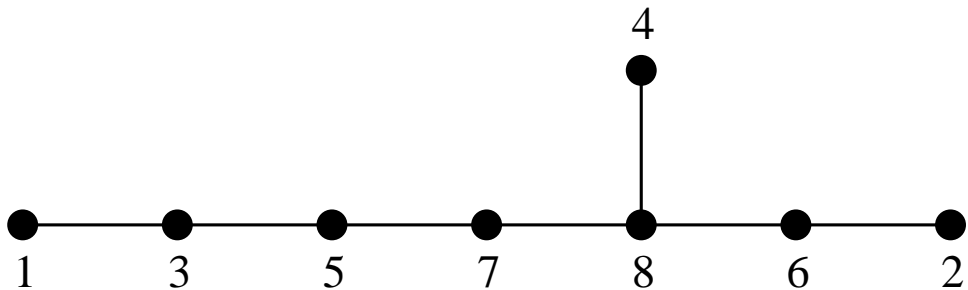}}
\end{figure}

In the massive limit ($\mu\rightarrow\infty$), the ratios 
\begin{equation}
R_{j}\; =\;\frac{F_{j+1}}{F_{1}}
\label{e:rj}
\end{equation} 
approach the particle mass ratios of the corresponding massive 
field theory. The masses of the eight stable particles are
proportional to 
the entries of the Perron-Frobenius eigenvector of the 
Cartan matrix of the Lie algebra $\mbox{E}_8$ and
their ratios (ordered by magnitude) are given by
\begin{equation}
\begin{array}{rclcl}
m_1/m & = & 1 & & \\*
m_2/m & = & 2\,\cos(\pi/5) & = & 1.618\, 034\ldots \\*
m_3/m & = & 2\,\cos(\pi/30) & = & 1.989\, 044\ldots \\*
m_4/m & = & 4\,\cos(\pi/5)\,\cos(7\pi/30) & = & 2.404\, 867\ldots \\*
m_5/m & = & 4\,\cos(\pi/5)\,\cos(2\pi/15) & = & 2.956\, 295\ldots \\*
m_6/m & = & 4\,\cos(\pi/5)\,\cos(\pi/30) & = & 3.218\, 340\ldots \\*
m_7/m & = & 8\,\cos^2(\pi/5)\,\cos(7\pi/30) & = & 3.891\, 157\ldots \\*
m_8/m & = & 8\,\cos^2(\pi/5)\,\cos(2\pi/15) & = & 4.783\, 386\ldots
\end{array}
\label{e:masses}
\end{equation}
where $m=m_1$ defines the mass scale. The Dynkin diagram of 
$\mbox{E}_8$ is shown in Fig.~\ref{f:dd}, where our labeling
of nodes follows the ordering of the masses.

In the TBA calculation of Ref.~\cite{BNW94}, each of the corresponding
eight massive particles is associated to a particular string in the 
BA solution. In addition, it is assumed that, apart
from these eight strings and the one-strings forming 
the vacuum solutions, no other string types occur in 
thermodynamically relevant quantities. With these 
assumptions, the TBA equations imply that the 
hole-type excitations in the one-strings have vanishing density.
The density of one-strings can be eliminated from 
the calculation, which finally leads to the $8\times 8$ scattering 
matrix of the $\mbox{E}_8$ factorized scattering theory \cite{BNW94}.

In Table~\ref{t:st}, the nine string types are given, labeled
by $t=0,1,\ldots 8$ in the order used in Ref.~\cite{BNW94},
which corresponds to the usual labeling of the $\mbox{E}_8$ 
Dynkin diagram rather than that used in Fig.~\ref{f:dd}. 
{}From the data of Table~\ref{t:st}, the string of type $t$
consists of $n^{(t)}$ roots of the form
\begin{equation}
v^{(t)}_k = v^{(t)} + \frac{i\pi}{32}
\left(\Delta^{(t)}_k + 16\,\varepsilon^{(t)}_k\right)
\label{e:strings}
\end{equation}
where $v^{(t)}$ denotes the center of the string on the real line.

\section{Numerical Results}
\label{s:nr}

For the numerical treatment of the BAE (\ref{e:BAE}) we used a
modified Newton method \cite{Stoer}. The calculations were performed
in extended precision FORTRAN 
(with 16 byte real numbers) on an IBM workstation. 

As a first step, we solved the BAE at criticality ($q=0$)
for small system size and compared the corresponding 
eigenvalues (\ref{e:eval}) 
to those obtained by direct diagonalization of the transfer matrix
using high precision arithmetics. 
This also served as a test of the performance of our programs.
In particular, the {\em complete\/} set of solutions corresponding to 
zero-momentum eigenstates of the transfer matrix for systems of 
size $N\le 6$ was obtained.
In this way, the structure of the solutions for the largest eigenvalues 
could be identified. These solutions were followed as a function of 
the elliptic nome $q$. The results of this analysis were then 
generalized to larger system size by looking for solutions of the
same type which differ by additional real roots only.

Since we are mainly interested in the mass ratios in the scaling
limit, we exclusively considered zero momentum states. For solutions
of the BAE, the momentum is obtained from the eigenvalue $\Lambda(0)$
(\ref{e:eval}) at spectral parameter $u=0$, where the transfer matrix
reduces to a shift operator.

Because we want to follow solutions for varying $q$, which means 
that we have to repeat the calculation for a large number of values
of $q$, we limited ourselves to systems of size $N\le 100$.
However, in certain cases the numerical accuracy of the calculations 
becomes the major problem. This happens for instance for BA solutions 
which contain roots that are very close to singularities of the BAE,
or for solutions where differences between roots 
(respectively their real parts) become extremely small.
These situations typically show up if one considers large values of the
elliptic nome $q$ (where the meaning of ``large'' depends on the
system size), but also for the massive excitations of the
critical dilute $\mbox{A}_3$ model discussed in Section~\ref{ss:me} 
below. As far as we can see, there are only two ways to go beyond
the limitations imposed by these numerical problems; either by
using higher precision arithmetics, or, to a lesser extent, 
by adjusting the program according to each specific type of 
BA solution (which we did to
some extent by treating the ubiquitous complex conjugate pairs
of BA roots with imaginary parts close to $\pm 11\pi/32$ in a 
special way).

\subsection{Critical Conformal Spectrum}
\label{ss:ccs}

\begin{figure}
\caption{Low-energy part of conformal zero-momentum spectrum for 
the critical dilute $\mbox{A}_3$ model.}
\label{f:conf}\nopagebreak
\centerline{\epsfxsize=0.9\columnwidth \epsfbox{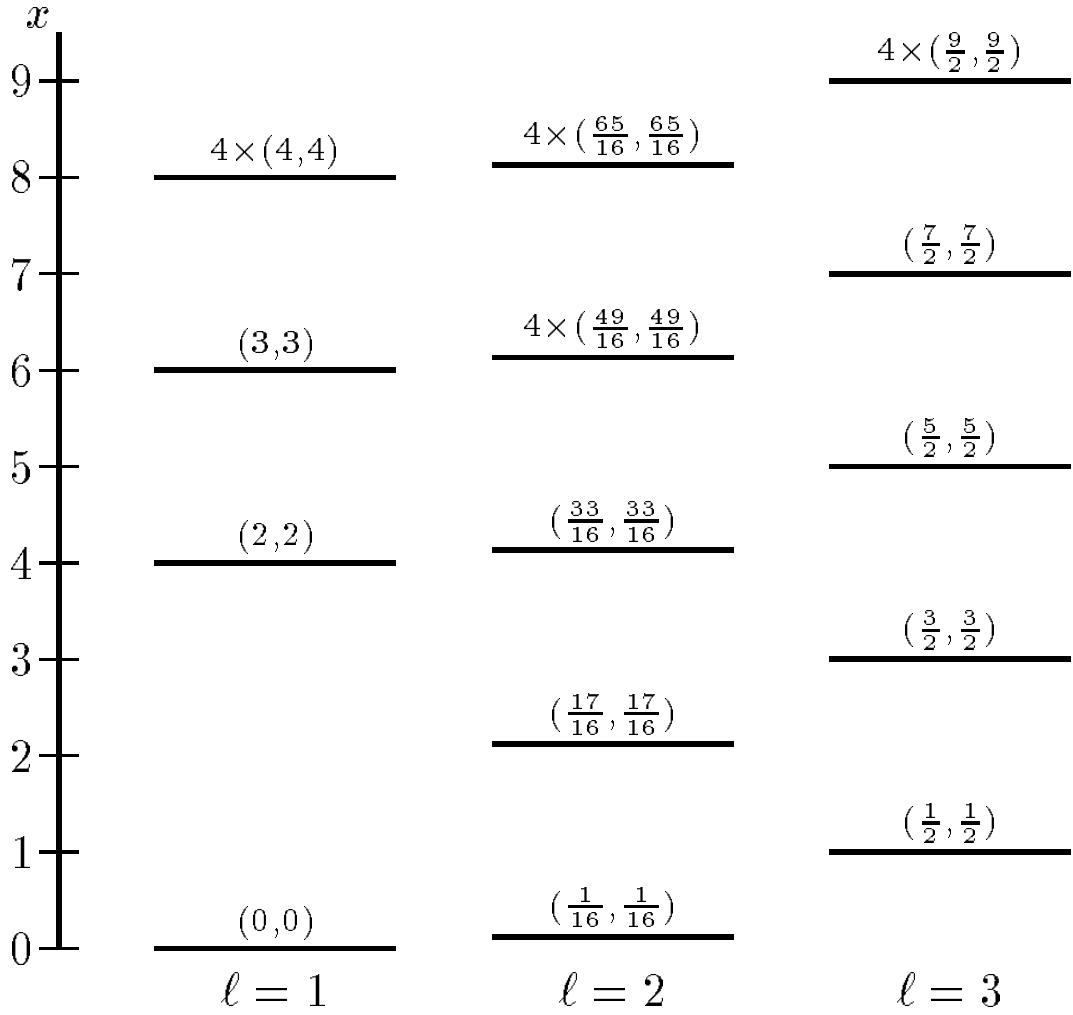}}
\end{figure}

In the critical limit ($q=0$, $N\rightarrow\infty$),
the largest eigenvalues of the transfer matrix organize
according to the characters of representations of the 
Virasoro algebra with central charge
$c=1/2$ as described in Section~\ref{ss:cs} above. Schematically,
the resulting values for the scaling dimensions $x$ in the
zero-momentum sector and their distribution into
the three sectors labeled by $\ell=1,2,3$ 
are shown in Fig.~\ref{f:conf}.

In Table~\ref{t:c}, numerical values for the central
charge and the lowest scaling dimensions obtained from
the corresponding solutions of the BAE for systems of
different sizes are presented. Here, Eq.~(\ref{e:cc}) with
the exact value of the bulk free energy \cite{WBN92,War93} 
of the dilute $\mbox{A}_3$ model was used. 
All zero-momentum states with a critical scaling dimension
of $x\le 7$ have been found, compare Fig.~\ref{f:conf},
except one of the four excitations
with conformal dimensions $(1/16+3,1/16+3)$ for which, despite
some effort, we have not been able to find the corresponding 
solution of the BAE. In particular, this includes all cases 
without degeneracy, i.e., where only a single zero-momentum
state of a certain scaling dimension occurs in the conformal tower.

To keep the three remaining BA solutions with scaling 
dimension $x_9=6+1/8$ apart, we distinguish them by 
subscripts $a$, $b$, and $c$. Actually, it turns out that
the two solutions denoted by $b$ and $c$ yield the same 
eigenvalue even for finite systems, which remains true
also off-criticality. Thus, we do not need to consider them
separately unless we are interested in the BA solutions,
which of course are different, see Section~\ref{ss:bar} below.

While the numerical data are in perfect agreement with the exact
values, it should be noted that the string structure of the
critical BA solutions have no apparent similarity to those
proposed in Ref.~\cite{BNW94}. We shall come back to this point later.

\subsection{Massive Excitations at Criticality}
\label{ss:me}

Analyzing the spectrum of largest eigenvalues of the transfer
matrix, one observes that, in addition to the conformal
spectrum discussed above, the spectrum contains {\em massive\/} 
excitations. More precisely, these are states with eigenvalues
$\Lambda^{(0)}_k$ for which the quantities
\begin{equation}
y_k\; = \;\log(\Lambda^{(0)}_0/\Lambda^{(0)}_k) 
\end{equation}
converge to a non-zero limit (mass) as $N\rightarrow\infty$,
compare Eq.~(\ref{e:xj}) for the conformal states where
this limit gives zero.

Apparently, a number of different masses is involved,
which are again characterized by particular
strings in the BA solutions. Strikingly, the strings which we observed are
among those listed in Table~\ref{t:st}. We have found explicit
solutions containing strings of type $t=1,2,3,4$, and $7$.
In all cases, the roots forming these strings are --- even 
for small systems --- located extremely close to the singularity of
the BAE which they approach for $N\rightarrow\infty$. This asks for
a careful numerical treatment, and severely limits the system sizes
we can treat. With the numerical accuracy of our program, we
were able to get reliable results for systems of at most 
20--30 sites, depending on the particular solution.

An example is given in Table~\ref{t:me}. Here, the solutions
obtained by adding a number of ``massive'' two-strings ($t=1$) 
to the ground-state solution of a system of size $N=4$ are
presented. {}From the eigenvalues for larger systems, one
can see that each additional ``massive'' two-string just adds
the same mass, thus substantiating our interpretation in terms
of massive particles.

Apparently, it is possible to add arbitrary numbers
of these ``massive'' strings not only to the ground-state,
but to {\em each\/} solution corresponding to a conformal state. 
This means that the spectrum (at the critical point $q=0$)
contains infinitely many copies 
of the complete conformal spectrum,
shifted with respect to the ground-state by a mass determined
by the collection of ``massive'' strings.

As to the observed ``massive'' strings, they fall into two classes:
for $t=1,3$, and $7$, we have a single solution of the BAE,
whereas for $t=2$ and $4$ the solutions belong to eigenvalues
which are doubly degenerate (for finite systems). Labeling the
masses associated to the ``massive'' string type $t$ by
$M_t$, we obtain the numerical estimates
\begin{equation}
\begin{array}{rcl@{\qquad\quad}rcl}
M_7/M_1 & \approx & 1.85186 & M_2/M_1 & \approx & 1.78 \\
M_3/M_1 & \approx & 3.17213 & M_4/M_1 & \approx & 3.78
\end{array}
\end{equation}
for the ratios of the masses with respect to the smallest mass
which is associated to the two-string ($t=1$). Note that these
numbers are calculated from the eigenvalues of the transfer
matrix with spectral parameter $u=3\lambda/2$. 

This situation is similar to what is observed in the Hubbard
model, where one has two types of excitations: so-called ``holons'' 
(carrying charge but no spin) and ``spinons'' (carrying spin but no 
charge) \cite{KSZ90,EK93}. 
Generically, both excitations are massless, and the
corresponding field theory consists of two coupled $c=1$ CFT.
However, if the filling fraction (number of electrons per lattice site)
is chosen to be precisely one, the holons become massive while the
spinons stay massless. At low energies, this theory is then
described by a $c=1$ CFT with additional massive 
excitations \cite{kareljan}.

\subsection{Bethe Ansatz Roots Off-Criticality}
\label{ss:bar}

As mentioned above, the string structure of the
BA roots at criticality does not seem to agree with
the predictions of Ref.~\cite{BNW94}. However, if one
follows a particular solution as a function of the nome $q$,
one finds that in many cases one encounters singularities of the
BAE where the string structure changes. 

\begin{figure}
\caption{Solution of BAE for the second largest eigenvalue for
varying nome $q$. Thick lines denote the real parts
(normalized by $-\log(q)$), thin lines the imaginary parts 
(in units of $\pi$) of the BA roots.}
\label{f:bas}\nopagebreak
\centerline{\epsfxsize=\columnwidth \epsfbox{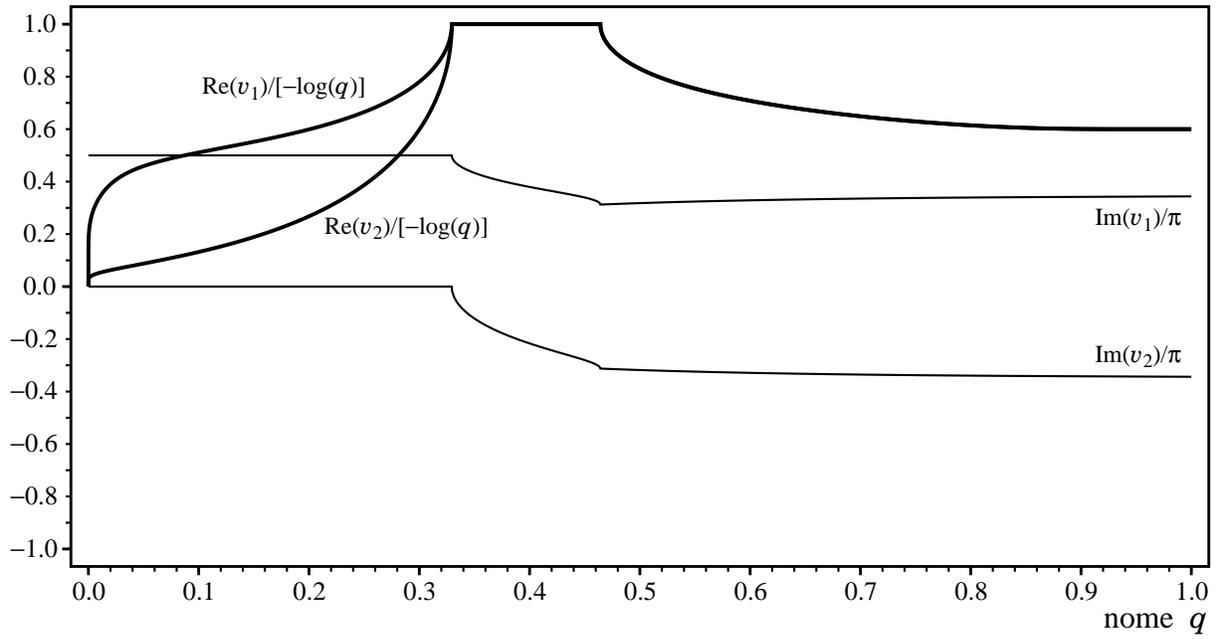}}
\end{figure}

The simplest example is given by the second
largest eigenvalue, which corresponds to the conformal
spin density field with dimensions $(1/16,1/16)$ in the critical
limit and to the lightest massive particle in the massive
scaling limit. At criticality,
the corresponding solution of the BAE differs from that
of the largest eigenvalue (which contains real roots $v_j$ only)
by a single root with imaginary part $\pi/2$. However, as can
been seen from Fig.~\ref{f:bas}, as $q$ is increased from $q=0$,
the complex root and one real root approach each other, until
their real parts agree. At that point, the real parts of the two
roots stay the same, but they move in the imaginary direction
until they form a complex conjugate pair of roots (hence a two-string),
which then persists for large values of $q$. 

Comparing systems of different size, this mechanism stays always 
the same, the additional real roots just play the r\^{o}le of spectators. 
Moreover, the transition from one string type to another takes
place at approximately the same value of the scaling parameter
$\mu$ (\ref{e:mu}) as the system size is increased. More precisely,
the scaling parameter values approach a non-zero limit as 
the system size tends to infinity. This implies that
although $q\rightarrow 0$ in the scaling limit, for large values
of $\mu$ (and in particular for the massive scaling limit 
$\mu\rightarrow\infty$),
the two-string is the relevant type of solution.

The same scenario applies to the other excitations we considered.
In Table~\ref{t:ss}, we compiled numerical values of the non-real
BA roots for systems of size $N=100$, both at criticality 
($q=0$, i.e.\ $\mu=0$) and for a rather large value $\mu\approx 39.4$ 
of the scaling parameter (corresponding to $q=7/1000$ for $N=100$).
Note that for $q\ne 0$ the real part of the roots $v_j$ 
can be shifted by multiples of $-i\pi\tau=-\log(q)\in\mathbb{R}_{+}$
($0<q<1$) as discussed in Section~\ref{ss:bae} above. 
Using this property, we arranged the solutions such that
the values of the non-real roots have a real
part in the interval $[0,-\log(q))$. Actually, 
as Table~\ref{t:ss} suggests, all non-real roots are located
in the vicinity of $-\log(q)/2$ for sufficiently large scaling 
parameters $\mu$. This also distinguishes these solutions from
the ``massive'' strings at criticality considered above, whose
centers are located close to the origin.
Furthermore, the values of the nome $q$
where changes in the string patterns for $N=100$ occur are
also indicated in Table~\ref{t:ss}. All five single-particle
states show at least one such point, whereas some of the 
two-particle states do not change as $q$ is varied. 

Although the relation between the scaling exponents on one
end and the content of massive particles on the other is determined
by the field theory, we are not aware that it has been calculated.
Our results given in Table~\ref{t:ss} give an unambiguous connection for 
the lowest states, but they do not suggest an obvious pattern that 
can be generalized to the higher excitations.

Comparing with the proposed string types of Ref.~\cite{BNW94}
given in Table~\ref{t:st} and by Eq.~(\ref{e:strings}),
one finds that the solutions for
large scaling parameter agree with the predictions
apart from the two states which contain a particle of mass $m_4$.
Here, in place of a five-string with imaginary parts
$\pm\pi/4$, $\pm 3\pi/8$ and $\pi/2$, the solution found
numerically on first view looks more like a seven-string with imaginary
parts $\pm\pi/32$, $\pm 7\pi/32$, $\pm 13\pi/32$ and $\pi/2$.
However, in comparison to the other single-particle states,
the real parts of the roots forming this ``string'' are quite
far apart, which might be an indication for another change of
pattern occurring at a larger value of $q$, which of course
we cannot exclude on the basis of our numerical data.

\begin{figure}
\caption{Solution of BAE corresponding to the single-particle
state with mass $m_4$ in the scaling limit. The system size is
$N=10$. Thick lines denote the real parts
(normalized by $-\log(q)$), thin lines the imaginary parts 
(in units of $\pi$) of the BA roots $v_j$. For each real part,
the numbers in square brackets give the (approximate) imaginary 
parts (in units of $\pi/32$) for large values of the scaling 
parameter $\mu$.}
\label{f:bam4_10}\nopagebreak
\centerline{\epsfxsize=\columnwidth \epsfbox{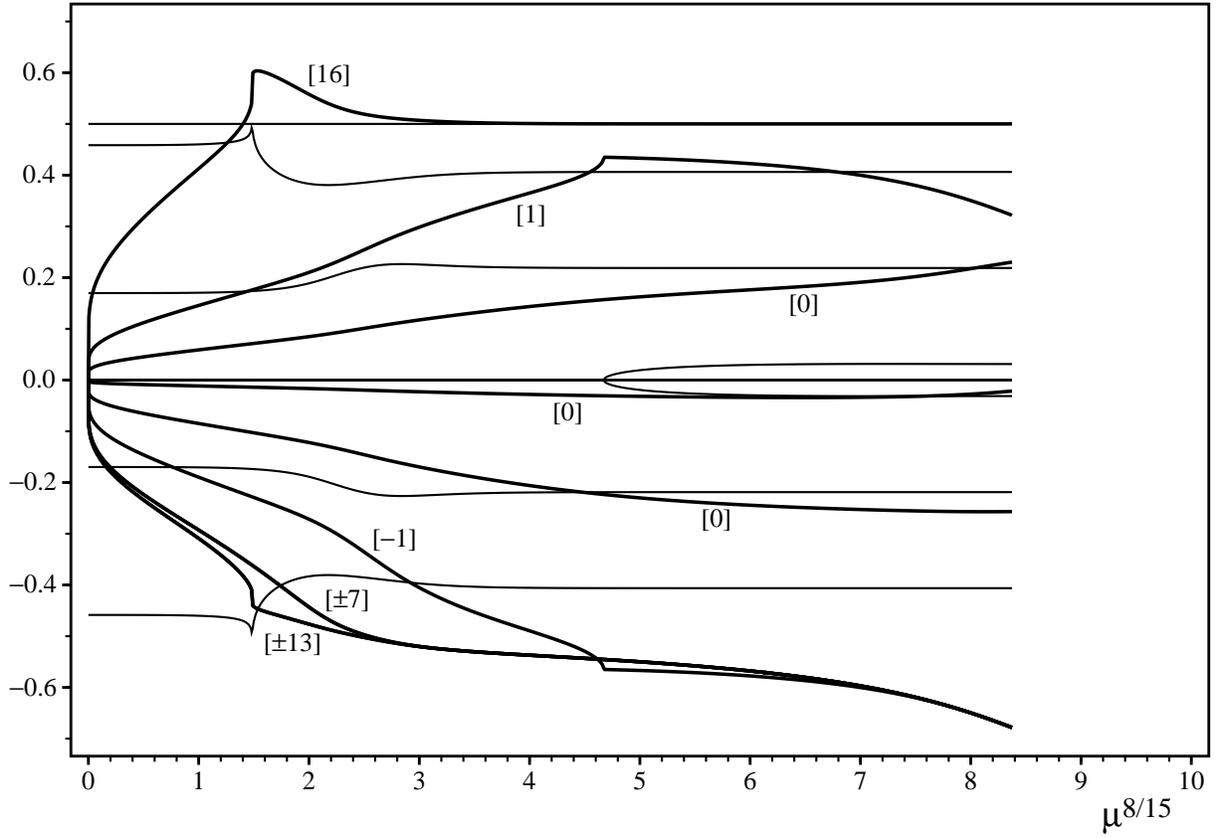}}
\end{figure}

\begin{figure}
\caption{Same as Fig.~\protect\ref{f:bam4_10}, but for
system size $N=20$. For clarity, the numbers in square brackets have been
omitted for all real roots.}
\label{f:bam4_20}\nopagebreak
\centerline{\epsfxsize=\columnwidth \epsfbox{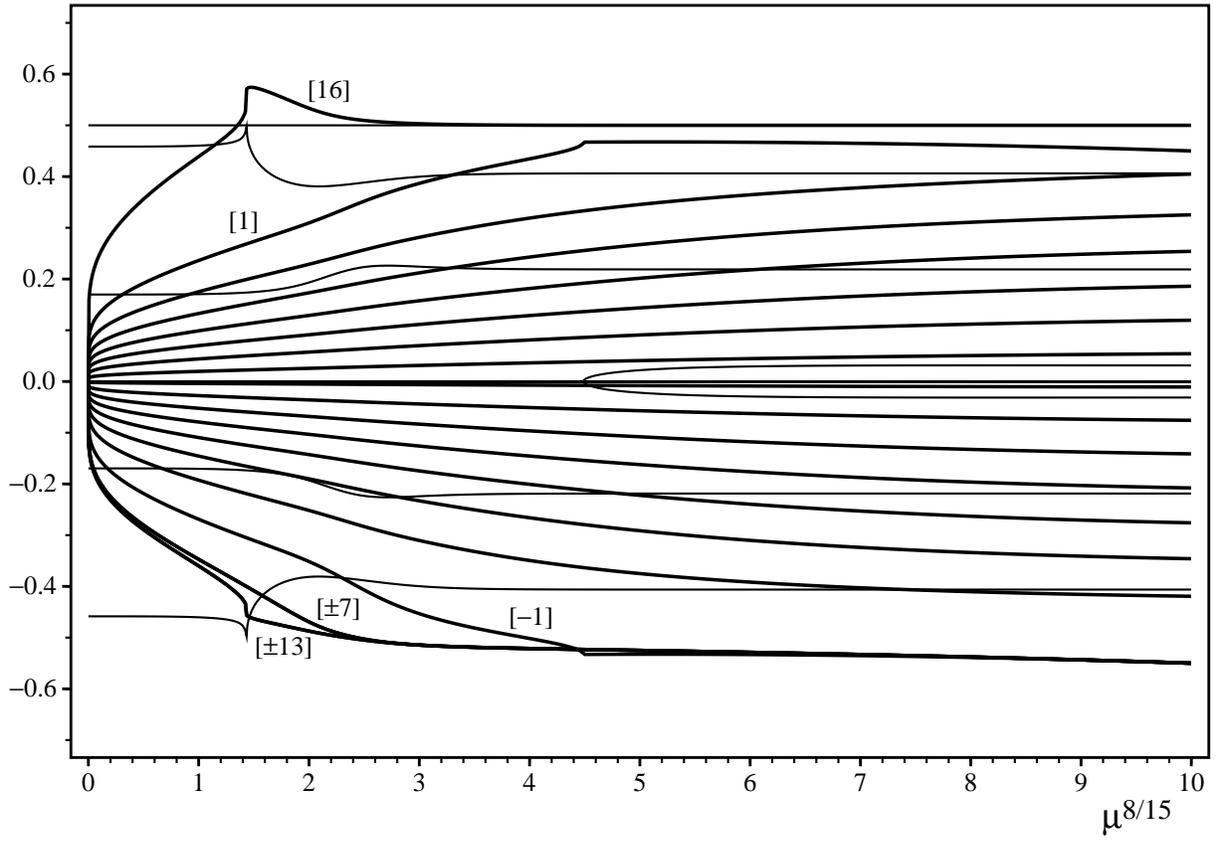}}
\end{figure}

Let us have a closer look at the single particle state with mass
$m_4$. Figures~\ref{f:bam4_10} and \ref{f:bam4_20} show the BA roots 
of the corresponding solutions for systems with $N=10$ and
$N=20$ sites, respectively. For convenience, we plotted the
real and imaginary parts of the roots, normalized by $-\log(q)$ and
by $\pi$ as in Fig.~\ref{f:bas}, against $\mu^{8/15}=q^{8/15}N$.
There are two obvious points where non-analyticities can be
seen in the imaginary parts, which both are situated at approximately 
the same value of the scaling parameter for the two different systems.
However, Fig.~\ref{f:bam4_10} may suggest that another change will
occur at a larger value of $\mu$ than numerically accessible
by our routines, as the real parts of one real root and one of the
two-strings approach each other. But, comparing with the same region of
Fig.~\ref{f:bam4_20}, no indication of this behavior remains,
and the same holds true for the larger systems (up to $N=100$) 
which we examined. {}From this, we conclude that either there
is no further change in the string pattern, or it has to occur for
a very large value of the scaling parameter $\mu$, which however
seems rather unlikely to us.

Though this result is somewhat inconclusive, let us examine the
string pattern of this solution in more detail. In particular,
we are interested in the question whether there are holes in the
one-strings in this case. To see this, we consider the
``phase function''
\begin{eqnarray}
\varphi(v)& =&\frac{1}{2\pi i}\,
\log\left[-\,\omega^{-\ell}\:
{\left(\frac{\vartheta_{1}(iv+\lambda)}
{\vartheta_{1}(iv-\lambda)}\right)}^{N}\right.\nonumber \\*
& & \;\;\:\cdot\,\left.\prod_{k=1}^{N} 
\frac{\vartheta_{1}(iv-iv_k-2\lambda)\,\vartheta_{1}(iv-iv_k+\lambda)}
{\vartheta_{1}(iv-iv_k+2\lambda)\,\vartheta_{1}(iv-iv_k-\lambda)}
\right] \label{e:ph}
\end{eqnarray}
which is basically the logarithm of the BAE (\ref{e:BAE}), but now
$v_k$ denote the roots of our particular BA solution, and we consider
$\varphi$ as a function of the complex variable $v$.
By definition, $\varphi(v_j)\; \in\;\mathbb{Z}$
for all the roots $v_j$, $j=1,\ldots,N$. Restricting to 
real values of $v$, $\varphi(v)$ therefore takes integer values at 
all real solution $v=v_j\in\mathbb{R}$, hence for all one-strings.

\begin{figure}
\caption{The phase function $\varphi(v)$ (\protect\ref{e:ph})
for the solution of BAE corresponding 
to the single-particle state with mass $m_4$ in the scaling limit. 
The system size is $N=20$, $q=1/5$. Horizontal lines are drawn at 
integer values, crosses at intersections with the graph of 
$\varphi(v)$ denote the real roots, and
vertical lines indicate the positions of the real parts of
the other seven roots.}
\label{f:m4ph}\nopagebreak
\centerline{\epsfxsize=\columnwidth \epsfbox{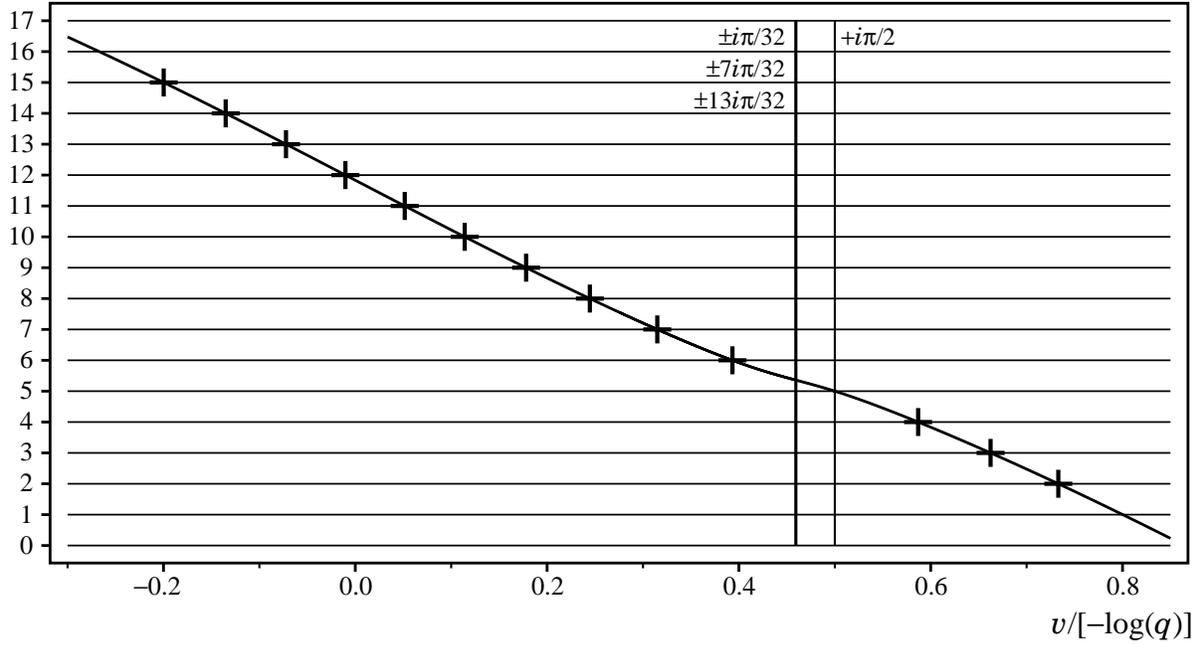}}
\end{figure}

Fig.~\ref{f:m4ph} shows the $\varphi(v)$ for the solution
under consideration. Here, the size of the system is $N=20$,
and $q=1/5$. The horizontal lines are drawn at integer values
(the precise numbers are not relevant as they depend on the
choice of branch in Eq.~(\ref{e:ph})),
and the crosses located on intersections of these lines
with the graph of $\varphi(v)$ denote the position of the real roots.
The vertical lines are placed at the real parts of the remaining
seven non-real roots. As observed above, the single root with
imaginary part $\pi/2$ lies somewhat separated from the three
complex conjugate pairs which are very close together (such that the
three vertical lines appear as one thicker line in the figure).
Moreover, the location of this single root coincides {\em precisely\/}
with a hole in the one-strings, which may seem to disagree with  
the findings of Ref.~\cite{BNW94}. However, the behavior of the roots as 
a function of parameters like $q$ suggests that the roots 
with imaginary part $\pi/2$ are simply an alternative locus of
the one-string.

On the basis of the numerical data, it thus seems that the
string structure associated to the mass $m_4$ consists of the rather
complicated pattern of a seven-string involving three complex conjugate 
pairs of roots (with imaginary parts
$\pm\pi/32$, $\pm 7\pi/32$ and $\pm 13\pi/32$) and a neighboring
root at $i\pi/2$ which comes together with a hole in the one-string 
distribution. Of course, in the infinite size limit the distance
between the real parts of these roots becomes infinitesimal.
It is not obvious what happens in the TBA calculation of
Ref.~\cite{BNW94} if their string type $t=8$ (see Table~\ref{t:st})
is replaced by the observed pattern, especially as we did not
see the strings associated to the three largest masses. 
Nevertheless, we believe that the main
ideas underlying the TBA treatment in Ref.~\cite{BNW94} are correct,
and that only details of the calculation would be affected.

Before we move on to the discussion of the scaling function,
two short remarks regarding the BA solutions are in order.
It should be noted that the ``massive'' strings 
mentioned previously in Section~\ref{ss:me},
which yield the massive excitations {\em at criticality\/},  
do not undergo similar changes.
These strings are always located very close to singularities
of the BAE, and move even closer as the nome $q$ is increased.
Moreover, their centers lie at the origin,
see e.g.\ Table~\ref{t:me}, whereas for the states belonging
to the $\mbox{E}_8$ integrable field theory the strings
cluster around the value $-\log(q)/2$.
Finally, let us mention that a similar
behavior of BA roots has recently been observed in the
solution of the XXZ Heisenberg chain, where the boundary twist 
acts as the varying  parameter~\cite{FumItoOota}.

\subsection{Scaling Functions and Mass Ratios}
\label{ss:sf}

Let us now turn to the results for the scaling functions
$F_0(\mu)$ (\ref{e:f0}) and $F_j(\mu)$ (\ref{e:fj}).
For the latter, all zero-momentum states with a critical 
scaling dimension of $x\le 7$ are considered,
except the one missing excitation
with conformal dimensions $(1/16+3,1/16+3)$. 
In all cases, we use data from systems
of size $N=50$, $N=75$ and $N=100$. For these sizes and the 
values of $q$ we considered, it turns out that
the corrections to scaling are so small that one can hardly 
recognize them in our figures.

\begin{figure}
\caption{Scaling function $F_0(\mu)$ (\protect\ref{e:f0})
for the largest eigenvalue obtained
from systems of size $N=50$, $75$, and $100$. Here, the logarithm
of $F_0(\mu)$ is plotted against 
$\mu^{8/15}=q^{8/15}N$ (\protect\ref{e:mu})}
\label{f:gsf}\nopagebreak
\centerline{\epsfxsize=\columnwidth \epsfbox{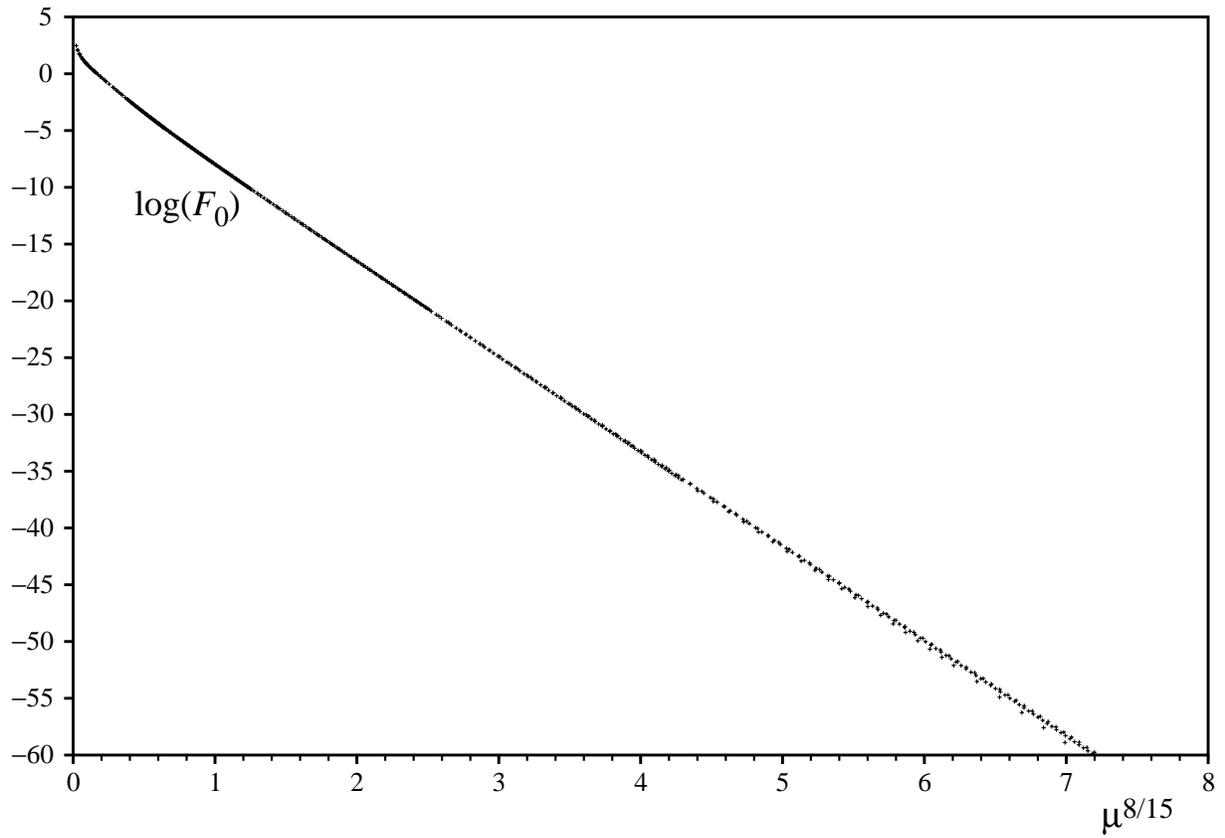}}
\end{figure}

In Fig.~\ref{f:gsf}, the (natural) logarithm of $F_0(\mu)$ 
(which as defined in Eq.~(\ref{e:f0}) is positive) is shown
as a function of $\mu^{8/15}=q^{8/15}N$ (\ref{e:mu}).
Apart from the behavior for small $q$, the plot is
nicely linear, showing the exponential decrease of $F_0$
as a function of $\mu^{8/15}$ down to about $\exp(-60)$
where the difference between the eigenvalues for the finite sizes
and the bulk limit value becomes smaller than our numerical precision,
which also proves the performance of our numerical routines.
Small deviations from scaling can be seen for larger values of $q$.

\begin{figure}
\caption{Scaling functions $F_j(\mu)$ (\protect\ref{e:fj})
for the excitations obtained from systems of size $N=100$.}
\label{f:sf}\nopagebreak
\centerline{\epsfxsize=\columnwidth \epsfbox{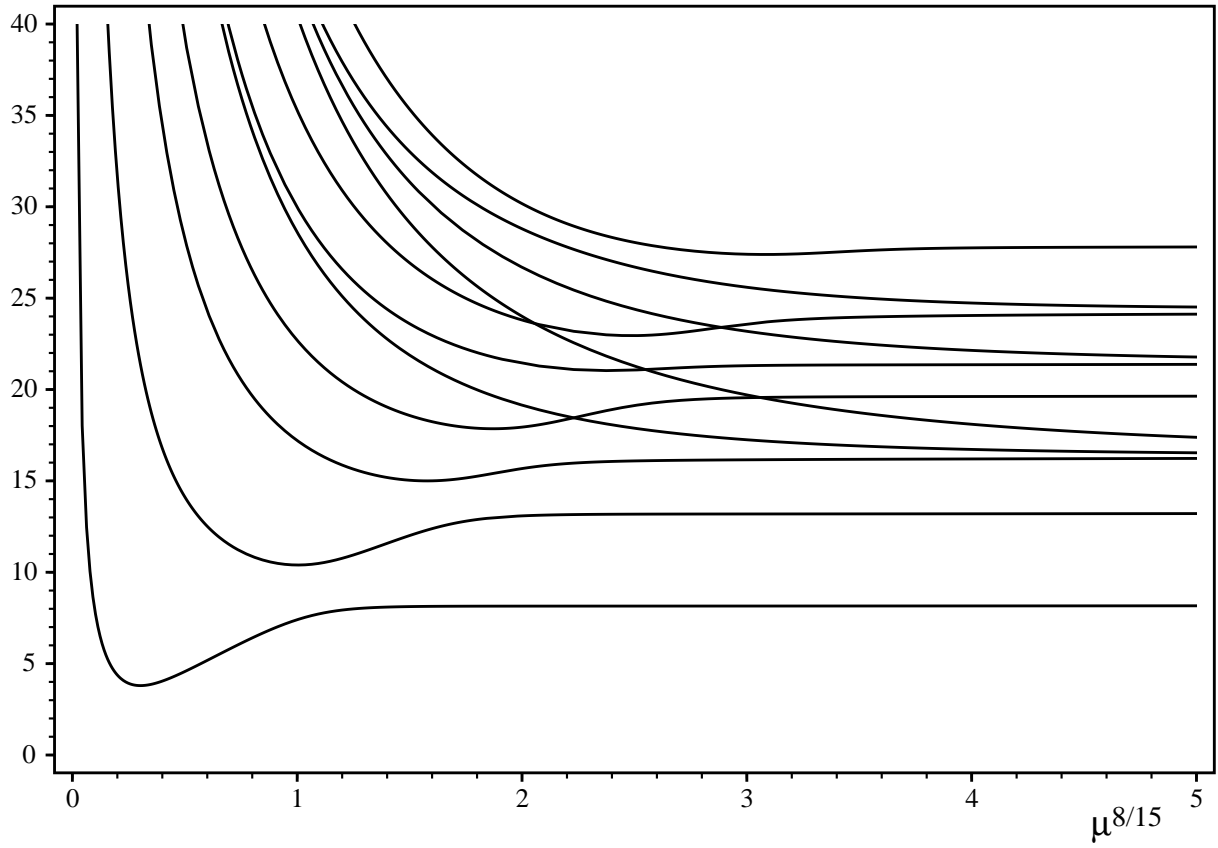}}
\end{figure}

\begin{figure}
\caption{Ratios of scaling functions $R_j(\mu)$ (\protect\ref{e:rj}).
The individual data points shown stem from systems of size 
$N=50$, $75$, and $100$. Subscripts $a,b,c$ distinguish
different solutions with identical critical scaling dimensions.
The $\mbox{E}_8$ mass ratios are indicated.}
\label{f:ratios}\nopagebreak
\centerline{\epsfxsize=\columnwidth \epsfbox{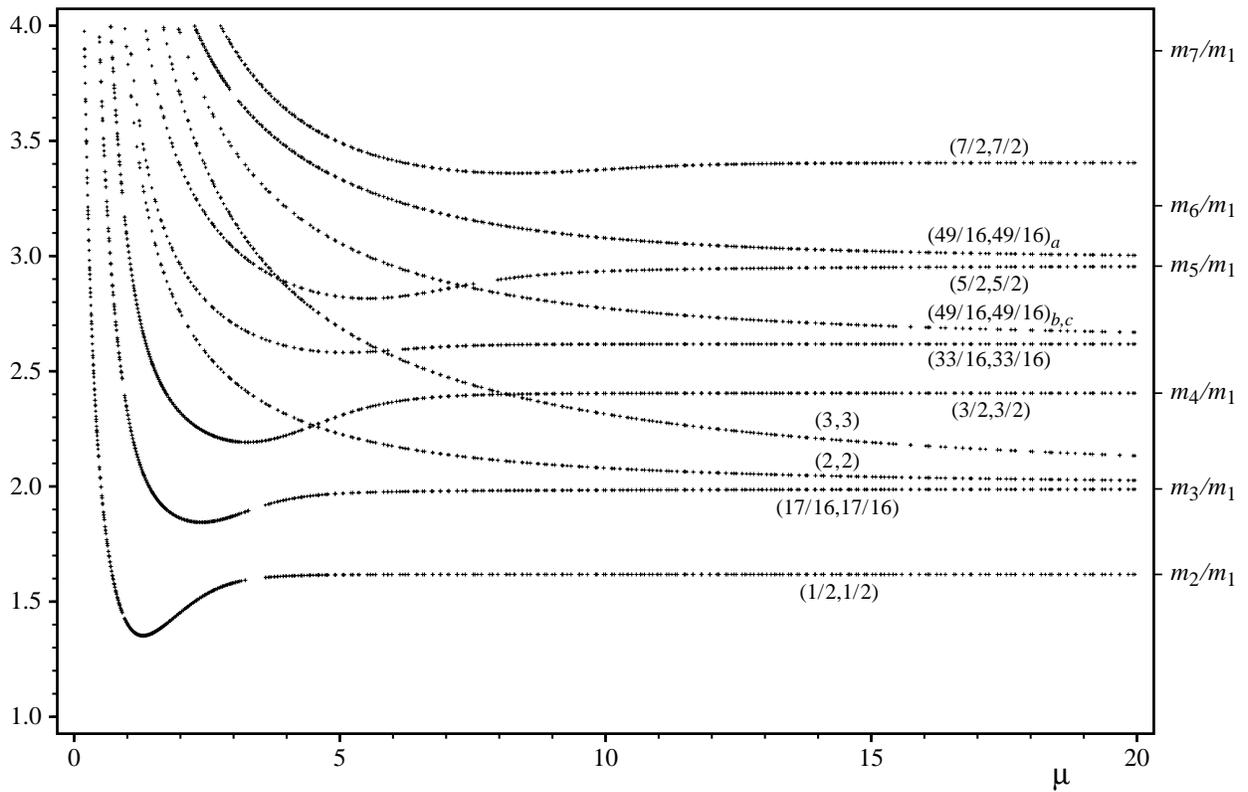}}
\end{figure}

The scaling functions for the excitations $F_j(\mu)$ (\ref{e:fj})
are displayed in Fig.~\ref{f:sf}. The curves shown are piecewise
linear plots connecting data points obtained from systems with $N=100$.
For convenience, we again used $\mu^{8/15}=q^{8/15}N$ (\ref{e:mu}) 
on the horizontal axis.
Qualitatively, the scaling functions agree 
with the results of previous numerical 
calculations for the Ising model in a magnetic 
field \cite{SagZamo,HenkSal,LauwRitt} and with the 
results of the truncated fermionic space approach, 
compare the figures given in Ref.~\cite{YurZam91}.
The one-particle states can be recognized by their characteristic
minima \cite{SagZamo}. This becomes clearer when we consider
their ratios $R_j(\mu)$ (\ref{e:rj}) which are presented in 
Fig.~\ref{f:ratios}. Here, we again show individual
data points which were obtained from systems of size
$N=50$, $N=75$, and $N=100$, which obviously were large enough to
keep corrections to scaling small. The ratios $R_j(\mu)$,
given here against the scaling variable $\mu$,
are labeled by the corresponding conformal dimensions
at criticality. We also indicated the single-particle mass ratios 
(\ref{e:masses}) of the $\mbox{E}_8$ field theory. Clearly,
the agreement is convincing, more detail on the approach
of the massive scaling limit is contained in Table~\ref{t:rat}.

\section{Conclusions}
\label{s:conc}

The spectrum of the dilute $\mbox{A}_3$ model has been studied by
numerical solution of the Bethe Ansatz equations, both at and 
off-criticality. 
This gives the correspondence between the lowest states of the 
conformal spectrum of the critical Ising model and those of the
massive $\mbox{E}_8$ field theory (note that the connection
given in Ref.~\cite{SagZamo} contains an obvious mistake).

At criticality, the spectrum consists for one part 
of massless excitations
described by $c=1/2$ minimal CFT as in the
critical Ising model, but in addition to these it contains 
massive excitations.
Obviously, the masses of these excitations are linked to the
appearance of particular strings in the Bethe Ansatz solutions,
which can be interpreted as massive particles. In our numerical
analysis, we have seen at least five such ``massive'' strings.
It is intriguing that all these strings (showing up in solutions
{\em at criticality\/}) are among the conjectured list of 
``thermodynamically significant string types'' in the TBA 
analysis \cite{BNW94} of the massive scaling limit. 
This is strongly suggestive of a connection between the 
Lie algebra $\mbox{E}_8$ and the masses of these excitations.

However, these are not the states which yield Zamolodchikov's 
$\mbox{E}_8$ field theory of the Ising model in a magnetic field.  
Those correspond to massless excitations at the critical point,
which develop a mass due to the existence of the symmetry-breaking 
field. At criticality, the string structure of these states does
not agree with the predictions of \cite{BNW94} --- to solve this puzzle
had been the original motivation for this work. Our numerical results 
suggest a scenario that may solve this apparent contradiction: 
the string type of the relevant excitations undergoes a number
of changes as the field is switched on, and for large systems
these reorganizations take place at particular values of the
scaling parameter $\mu$. Therefore, the string structure
entering the massive scaling limit $\mu\rightarrow\infty$
is that observed for large values of the elliptic nome $q$. 

Numerically, we have been able to identify the single particle states 
up to the fifth mass, and a number of two-particle states.
Clearly, our results for the scaling functions are in complete
agreement with both the analytic values of the mass ratios
and with earlier numerical work on the Ising model in a magnetic field.
As to the Bethe Ansatz solutions, apart from one exception, the
string structure for large field is that conjectured in \cite{BNW94}.
The one exception concerns the string type associated to the particle
of mass $m_4$. We found two states which contain this particle ---
one being the single particle state with conformal dimensions 
$(3/2,3/2)$, the other a two-particle state which also contains
the lightest particle and has scaling dimensions $(7/2,7/2)$ at criticality.
In both cases, the observed string structures agree, 
but they differ from 
the one proposed in \cite{BNW94}, compare Table~\ref{t:ss}. 
Of course, it cannot
be ruled out completely that another change of string type appears
at a larger value of the scaling parameter, but we found no trace 
of such a  behavior for scaling parameters up to $\mu\lesssim 80$.
This clearly demands further clarification, and maybe the investigation
of the dilute $\mbox{A}_4$ model can lead the way. 
In the scaling limit, this model is described by an $\mbox{E}_7$ 
theory of factorized scattering, but has so far eluded a TBA approach
analogous to that of \cite{BNW94}. In this case, we have been able to 
identify {\em all\/} seven single-particle states and observed
interesting string solutions, details will be published 
soon \cite{GN}.

A number of interesting questions arise in connection with
the massive excitations of the critical dilute $\mbox{A}_3$ model.
This is one property of the dilute $\mbox{A}_3$ model which
distinguishes it from the proper Ising model in a magnetic field,
since the critical Ising model does not show such excitations.
Moreover, we suppose that this phenomenon is not particular to the 
specific model and will show up in the other dilute models as well. It 
would be interesting to understand the physical nature of these 
excitations, and to obtain analytic predictions for the observed 
mass ratios. Also, their dispersion relations have not been studied 
because we concentrated on momentum zero states throughout this work.

\acknowledgements

U.G.\ gratefully acknowledges financial support of the 
Samenwerkingsverband FOM/SMC Mathematische Fysica during his stay in 
Amsterdam, where most of this work was done. The authors thank 
M.\ Baake, B.~M.\ \mbox{McCoy}, P.~A.\ Pearce, K.\ Schoutens and 
S.~O.\ Warnaar for interesting discussions and helpful comments.







\clearpage

\narrowtext

\narrowtext

\narrowtext

\narrowtext

\mediumtext

\mediumtext

\mediumtext

\mediumtext

\mediumtext

\mediumtext

\widetext




\clearpage

\narrowtext
\begin{table}
\caption{The nine thermodynamically significant string types of 
         Ref.~\protect\cite{BNW94} and the corresponding masses.}
\label{t:st}\nopagebreak
\begin{tabular}{ccrcl}
\multicolumn{1}{c}{mass} & 
\multicolumn{1}{c}{$t$} & 
\multicolumn{1}{c}{$n^{(t)}$} & 
\multicolumn{1}{c}{$\varepsilon^{(t)}$} & 
\multicolumn{1}{c}{$\Delta^{(t)}/5$\rule[-1.3ex]{0ex}{4ex}} \\
\hline
  0 & 0 &  1 & 0 & (0) \rule[0ex]{0ex}{2.7ex}\\
$m_1$ & 1 &  2 & 1 & (-1,1) \\
$m_2$ & 7 &  4 & 0 & (-3,-1,1,3) \\
$m_3$ & 2 &  4 & 0 & (-4,-2,2,4) \\
$m_4$ & 8 &  5 & 1 & (-12,-8,0,8,12)\\
$m_5$ & 3 &  6 & 1 & (-7,-5,-1,1,5,7) \\
$m_6$ & 6 &  7 & 1 & (-14,-6,-2,0,2,6,14) \\
$m_7$ & 4 &  8 & 0 & (-10,-8,-4,-2,2,4,8,10) \\
$m_8$ & 5 & 10 & 1 & (-13,-11,-7,-5,-1,1,5,7,11,13)\rule[-1.3ex]{0ex}{2ex}
\end{tabular}
\end{table}

\widetext
\squeezetable
\begin{table}
\caption{Finite-size approximants for central charge $c$ and
smallest scaling dimensions $x_{j}$.}
\label{t:c}\nopagebreak
\begin{tabular}{rllllllllllll}
\multicolumn{1}{c}{\rule[-1.3ex]{0ex}{4ex} $N$} &
\multicolumn{1}{c}{$c$} &
\multicolumn{1}{c}{$x_1$} &
\multicolumn{1}{c}{$x_2$} &
\multicolumn{1}{c}{$x_3$} &
\multicolumn{1}{c}{$x_4$} &
\multicolumn{1}{c}{$x_5$} &
\multicolumn{1}{c}{$x_6$} &
\multicolumn{1}{c}{$x_7$} &
\multicolumn{1}{c}{$x_8$} &
\multicolumn{1}{c}{$x_{9a}$} &
\multicolumn{1}{c}{$x_{9b,c}$} &
\multicolumn{1}{c}{$x_{10}$} \\
\hline 
\rule[0ex]{0ex}{2.7ex}
 10&0.499$\,$681&0.125$\,$080&1.014$\,$498&2.196&3.148&4.272&4.417
   &5.449&6.608&6.771&6.804&7.942\\
 20&0.499$\,$920&0.125$\,$020&1.003$\,$543&2.142&3.034&4.062&4.191
   &5.098&6.142&6.274&6.274&7.196\\
 50&0.499$\,$987&0.125$\,$003&1.000$\,$563&2.128&3.005&4.010&4.135
   &5.015&6.022&6.148&6.148&7.030\\
 75&0.499$\,$994&0.125$\,$001&1.000$\,$250&2.126&3.002&4.004&4.130
   &5.007&6.010&6.135&6.135&7.013\\
100&0.499$\,$997&0.125$\,$001&1.000$\,$141&2.126&3.001&4.002&4.128
   &5.004&6.005&6.131& 6.131&7.007
\rule[-1.3ex]{0ex}{2ex} \\ 
\hline
\multicolumn{1}{c}{\rule[-1.3ex]{0ex}{4ex} $\infty$} &
\multicolumn{1}{c}{$1/2$} &
\multicolumn{1}{c}{$1/8$} &
\multicolumn{1}{c}{$1$} &
\multicolumn{1}{c}{$\!\!\! 2\! +\! 1/8\!\!\!$} &
\multicolumn{1}{c}{$3$} &
\multicolumn{1}{c}{$4$} &
\multicolumn{1}{c}{$\!\!\! 4\! +\! 1/8\!\!\!$} &
\multicolumn{1}{c}{$5$} &
\multicolumn{1}{c}{$6$} &
\multicolumn{1}{c}{$\!\! 6\! +\! 1/8\!\!$} & 
\multicolumn{1}{c}{$\!\! 6\! +\! 1/8\!\!$} & 
\multicolumn{1}{c}{$7$} \\
\end{tabular}
\end{table}

\widetext
\squeezetable
\begin{table}
\caption{Example of Bethe Ansatz solutions 
at criticality in sector $\ell=1$ which differ
from the ground-state solution for system size $N=4$ by a number 
of ``massive'' two-strings. Here, $\sigma=i\pi/32$.}
\label{t:me}\nopagebreak
\begin{tabular}{r@{.}lr@{.}lr@{.}lr@{.}lr@{.}lr@{.}l}
\multicolumn{2}{c}{\rule[-1.3ex]{0ex}{4ex} $N=4$} &
\multicolumn{2}{c}{$N=6$} &
\multicolumn{2}{c}{$N=8$} &
\multicolumn{2}{c}{$N=10$} &
\multicolumn{2}{c}{$N=12$} &
\multicolumn{2}{c}{$N=14$} \\
\hline \rule[-1.3ex]{0ex}{4ex}
$ 1$&$977\, 471\, 499$ & $ 1$&$980\, 014\, 437$ & $ 1$&$982\, 321\, 219$ & 
$ 1$&$984\, 408\, 860$ & $ 1$&$986\, 312\, 526$ & $ 1$&$988\, 062\, 914$ \\ 
\rule[-1.3ex]{0ex}{4ex}
$ 0$&$575\, 364\, 766$ & $ 0$&$577\, 361\, 036$ & $ 0$&$579\, 176\, 248$ & 
$ 0$&$580\, 821\, 977$ & $ 0$&$582\, 324\, 931$ & $ 0$&$583\, 708\, 690$ \\ 
\rule[-1.3ex]{0ex}{4ex}
$-0$&$105\, 369\, 741$ & $-0$&$105\, 850\, 465$ & $-0$&$106\, 289\, 271$ & 
$-0$&$106\, 688\, 316$ & $-0$&$107\, 053\, 693$ & $-0$&$107\, 390\, 875$ \\ 
\rule[-1.3ex]{0ex}{4ex}
$-0$&$886\, 306\, 204$ & $-0$&$888\, 679\, 105$ & $-0$&$890\, 833\, 363$ & 
$-0$&$892\, 784\, 131$ & $-0$&$894\, 563\, 861$ & $-0$&$896\, 200\, 997$ \\ 
\multicolumn{2}{c}{\rule[0ex]{0ex}{2.7ex}} &
$-0$&$000\, 000\, 072\:\pm$ &
$ 0$&$020\, 936\, 097\:\pm$ & 
$ 0$&$035\, 854\, 325\:\pm$ &
$ 0$&$047\, 525\, 711\:\pm$ &
$ 0$&$057\, 153\, 974\:\pm$ \\
\multicolumn{2}{c}{\rule[-1.3ex]{0ex}{2ex}} &
$10$&$999\, 999\, 764\:\sigma$ &
$10$&$999\, 999\, 615\:\sigma$ & 
$10$&$999\, 999\, 489\:\sigma$ &
$10$&$999\, 999\, 373\:\sigma$ &
$10$&$999\, 999\, 263\:\sigma$ \\ 
\multicolumn{2}{c}{\rule[0ex]{0ex}{2.7ex}} &
\multicolumn{2}{c}{} &
$-0$&$020\, 936\, 294\:\pm$ &
$-0$&$000\, 000\, 132\:\pm$ &
$ 0$&$014\, 769\, 565\:\pm$ &
$ 0$&$026\, 242\, 969\:\pm$ \\
\multicolumn{2}{c}{\rule[-1.3ex]{0ex}{2ex}} &
\multicolumn{2}{c}{} &
$10$&$999\, 999\, 730\:\sigma$ &
$10$&$999\, 999\, 570\:\sigma$ &
$10$&$999\, 999\, 459\:\sigma$ &
$10$&$999\, 999\, 320\:\sigma$ \\
\multicolumn{2}{c}{\rule[0ex]{0ex}{2.7ex}} &
\multicolumn{2}{c}{} &
\multicolumn{2}{c}{} &
$-0$&$035\, 854\, 556\:\pm$ &
$-0$&$014\, 769\, 869\:\pm$ &
$ 0$&$000\, 000\, 171\:\pm$ \\
\multicolumn{2}{c}{\rule[-1.3ex]{0ex}{2ex}} &
\multicolumn{2}{c}{} &
\multicolumn{2}{c}{} &
$10$&$999\, 999\, 719\:\sigma$ &
$10$&$999\, 999\, 564\:\sigma$ &
$10$&$999\, 999\, 444\:\sigma$ \\
\multicolumn{2}{c}{\rule[0ex]{0ex}{2.7ex}} &
\multicolumn{2}{c}{} &
\multicolumn{2}{c}{} &
\multicolumn{2}{c}{} &
$-0$&$047\, 525\, 969\:\pm$ &
$-0$&$026\, 243\, 304\:\pm$ \\
\multicolumn{2}{c}{\rule[-1.3ex]{0ex}{2ex}} &
\multicolumn{2}{c}{} &
\multicolumn{2}{c}{} &
\multicolumn{2}{c}{} & 
$10$&$999\, 999\, 712\:\sigma$ &
$10$&$999\, 999\, 563\:\sigma$ \\
\multicolumn{2}{c}{\rule[0ex]{0ex}{2.7ex}} &
\multicolumn{2}{c}{} &
\multicolumn{2}{c}{} &
\multicolumn{2}{c}{} &
\multicolumn{2}{c}{} &
$-0$&$057\, 154\, 256\:\pm$ \\
\multicolumn{2}{c}{\rule[-1.3ex]{0ex}{2ex}} &
\multicolumn{2}{c}{} &
\multicolumn{2}{c}{} &
\multicolumn{2}{c}{} &
\multicolumn{2}{c}{} &
$10$&$999\, 999\, 706\:\sigma$
\end{tabular}
\end{table}

\widetext
\squeezetable
\begin{table}
\caption{Numerically observed string solutions 
for systems of size $N=100$ for nome $q=0$ ($\mu=0$)
and $q=7/1000$ ($\mu\approx 39.4$). Only the 
values of the non-real roots $v_{j}$ are given. Here,
$\varrho=-\log(q)=-i\pi\tau$ and $\sigma=i\pi/32$.
For those roots whose real part is very close to $\varrho/2$
or whose imaginary part is almost an integer times $\sigma$,
such that the number of digits given is not sufficient to
see the slight differences, the superscripts $+$ and $-$ indicate 
on which side the actual data lie. 
We also include ranges of $q$ values within which 
changes of the string patterns are observed. The three solutions
which correspond to a critical scaling dimension of $6+1/8$ are
labeled by subscripts $a$, $b$, and $c$, where the latter two
yield identical eigenvalues for finite systems.}
\label{t:ss}\nopagebreak
\begin{tabular}{r@{$\,$}c@{$\,$}lcr@{$\:$}c@{$\:$}r@{}c@{}l%
r@{}c@{}l@{$\:$}c@{$\:$}r@{}c@{}lr@{$\:$}c@{$\:$}l}
$\rule[-1.3ex]{0ex}{4ex}(\Delta+r$&$,$&$\overline{\Delta}+\overline{r})$ &
mass &
\multicolumn{5}{c}{non-real roots at $q=0$} &
\multicolumn{7}{c}{non-real roots at $q=7/1000$} &
\multicolumn{3}{c}{pattern change}\\
\hline
\rule[0ex]{0ex}{2.7ex}
$(0$&$,$&$0)$                                 & $0$       
&  &  & 
\rule[-1.3ex]{0ex}{2ex}\\
\rule[0ex]{0ex}{2.7ex}
$(\frac{1}{16}$&$,$&$\frac{1}{16})$           & $m_1$
& $5.277\, 011$&$+$&\multicolumn{3}{l}{$16\:\sigma$}
& $0$&$.$&$500\, 000^{+}\varrho$&$\pm$&$11$&$.$&$000\, 000^{-}\sigma$
& $0.000\, 138$ & -- & $0.000\, 139$ \\
& & & & & & & & & & & & & & & &  
$0.000\, 161$&--&$0.000\, 170$ 
\rule[-1.3ex]{0ex}{2ex}\\
\rule[0ex]{0ex}{2.7ex}
$(\frac{1}{2}$&$,$&$\frac{1}{2})$             & $m_2$
& $-4.123\, 856$&$\pm$&$7$&$.$&$024\, 170\;\sigma$
& $0$&$.$&$500\, 000^{-}\varrho$&$\pm$&$4$&$.$&$999\, 988\;\sigma$
& $0.000\, 560$ & -- & $0.000\, 640$ \\
& & & 
& $-4.052\, 796$&$+$&\multicolumn{3}{l}{$16\:\sigma$}
& $0$&$.$&$500\, 000^{-}\varrho$&$\pm$&$15$&$.$&$000\, 012\;\sigma$
& & &
\rule[-1.3ex]{0ex}{2ex}\\
\rule[0ex]{0ex}{2.7ex}
$(\frac{1}{16}+1$&$,$&$\frac{1}{16}+1)$       & $m_3$
& $-3.791\,098$&$+$&\multicolumn{3}{l}{$16\:\sigma$} 
& $0$&$.$&$500\, 000^{+}\varrho$&$\pm$&$12$&$.$&$011\, 549\;\sigma$
& $0.000\, 440$&--&$0.000\, 450$ \\
& & &
& $3.719\,808$&$\pm$&$11$&$.$&$045\, 724\:\sigma$
& $0$&$.$&$500\, 000^{+}\varrho$&$\pm$&$9$&$.$&$988\, 451\;\sigma$
& $0.000\, 580$&--&$0.000\, 640$ 
\rule[-1.3ex]{0ex}{2ex}\\
\rule[0ex]{0ex}{2.7ex}
$(\frac{1}{2}+1$&$,$&$\frac{1}{2}+1)$         & $m_4$
& $-3.517\, 907$&$\pm$&$14$&$.$&$671\, 118\:\sigma$
& $0$&$.$&$485\, 956\;\varrho$&$\pm$&$0$&$.$&$992\, 132\;\sigma$
& $0.000\, 340$&--&$0.000\, 350$ \\
& & &
& $-3.458\, 496$&$\pm$&$5$&$.$&$430\, 532\:\sigma$
& $0$&$.$&$486\, 904\;\varrho$&$\pm$&$7$&$.$&$000\, 002\;\sigma$
& $0.002\, 900$&--&$0.002\, 950$ \\
& & &
& $3.981\, 241$&$+$&\multicolumn{3}{l}{$16\:\sigma$}
& $0$&$.$&$486\, 904\;\varrho$&$\pm$&$12$&$.$&$999\, 998\;\sigma$
& & & \\
& & & & & & & & 
& $0$&$.$&$500\, 000^{+}\varrho$&$+$&\multicolumn{3}{l}{$16\:\sigma$} 
& & & 
\rule[-1.3ex]{0ex}{2ex}\\
\rule[0ex]{0ex}{2.7ex}
$(2$&$,$&$2)$                                 & $2\, m_1$
& $-3.245\, 499$&$\pm$&$10$&$.$&$982\, 954\;\sigma$
& $0$&$.$&$482\, 282\;\varrho$&$\pm$&$11$&$.$&$000\, 000^{-}\sigma$
& & &
\\
& & & 
& $3.237\, 651$&$\pm$&$10$&$.$&$937\, 395\;\sigma$
& $0$&$.$&$517\, 718\;\varrho$&$\pm$&$11$&$.$&$000\, 000^{-}\sigma$
& & &
\rule[-1.3ex]{0ex}{2ex}\\
\rule[0ex]{0ex}{2.7ex}
$(\frac{1}{16}+2$&$,$&$\frac{1}{16}+2)$       & $m_1+m_2$ 
& $-3.282\, 380$&$\pm$&$10$&$.$&$815\, 346\;\sigma$
& $0$&$.$&$500\, 000^{+}\varrho$&$\pm$&$11$&$.$&$000\, 000^{-}\sigma$
& $0.001\, 000$&--&$0.001\, 100$ \\
& & & 
& $3.304\, 607$&$\pm$&$10$&$.$&$953\, 718\;\sigma$
& $0$&$.$&$500\, 002\;\varrho$&$\pm$&$15$&$.$&$000\, 054\;\sigma$
& $0.001\, 410$&--&$0.001\, 420$ \\
& & &
& $4.238\, 695$&$+$&\multicolumn{3}{l}{$16\:\sigma$} 
& $0$&$.$&$500\, 002\;\varrho$&$\pm$&$4$&$.$&$999\, 946\;\sigma$
& & & 
\rule[-1.3ex]{0ex}{2ex}\\
\rule[0ex]{0ex}{2.7ex}
$(\frac{1}{2}+2$&$,$&$\frac{1}{2}+2)$         & $m_5$     
& $-3.356\, 336$&$\pm$&$12$&$.$&$564\, 943\;\sigma$
& $0$&$.$&$500\, 000^{-}\varrho$&$\pm$&$8$&$.$&$998\, 677\;\sigma$
& $0.000\, 800$&--&$0.000\, 900$ \\
& & & 
& $-3.326\, 604$&$\pm$&$8$&$.$&$981\, 386\;\sigma$
& $0$&$.$&$500\, 000^{-}\varrho$&$\pm$&$11$&$.$&$000\, 000^{-}\sigma$
& $0.001\, 275$&--&$0.001\, 475$ \\
& & & 
& $3.297\, 948$&$+$&\multicolumn{3}{l}{$16\:\sigma$} 
& $0$&$.$&$500\, 000^{-}\varrho$&$\pm$&$13$&$.$&$001\, 324\;\sigma$
& & & 
\rule[-1.3ex]{0ex}{2ex}\\
\rule[0ex]{0ex}{2.7ex}
$(3$&$,$&$3)$                                 & $2\, m_1$ 
& $-2.875\, 085$&$\pm$&$11$&$.$&$000\, 662\;\sigma$
& $0$&$.$&$455\, 576\,\varrho$&$\pm$&$11$&$.$&$000\, 000^{+}\sigma$
& & &
\\
& & &
& $2.875\, 191$&$\pm$&$11$&$.$&$000\, 554\;\sigma$
& $0$&$.$&$544\, 424\,\varrho$&$\pm$&$11$&$.$&$000\, 000^{+}\sigma$
& & &
\rule[-1.3ex]{0ex}{2ex}\\
\rule[0ex]{0ex}{2.7ex}
$(\frac{1}{16}+3$&$,$&$\frac{1}{16}+3)_{a}  $ & $m_1+m_3$ 
& $-2.890\, 111$&$\pm$&$11$&$.$&$001\, 019\;\sigma$
& $0$&$.$&$500\, 000^{-}\varrho$&$\pm$&$11$&$.$&$000\, 000^{+}\sigma$
& $0.000\, 430$&--&$0.000\, 440$ \\
& & &
& $2.889\, 954$&$\pm$&$11$&$.$&$000\, 849\;\sigma$
& $0$&$.$&$500\, 000^{+}\varrho$&$\pm$&$11$&$.$&$928\, 013\;\sigma$
& $0.000\, 520$&--&$0.000\, 550$ \\
& & & 
& $4.741\, 946$&$+$&\multicolumn{3}{l}{$16\:\sigma$} 
& $0$&$.$&$500\, 000^{+}\varrho$&$\pm$&$10$&$.$&$071\, 987\;\sigma$
& $0.003\, 750$&--&$0.004\, 100$
\rule[-1.3ex]{0ex}{2ex}\\
\rule[0ex]{0ex}{2.7ex}
$(\frac{1}{16}+3$&$,$&$\frac{1}{16}+3)_{b}  $ & $m_1+m_2$ 
& $-3.555\, 509$&$\pm$&$8$&$.$&$138\, 452\;\sigma$
& $0$&$.$&$471\, 689\;\varrho$&$\pm$&$11$&$.$&$000\, 000^{+}\sigma$
& $0.000\, 370$&--&$0.000\, 380$ \\
& & &
& $-3.317\, 682$&$+$&\multicolumn{3}{l}{$16\:\sigma$} 
& $0$&$.$&$517\, 536\;\varrho$&$\pm$&$5$&$.$&$000\, 047\;\sigma$
& & &
\\
& & &
& $2.889\, 954$&$\pm$&$11$&$.$&$000\, 849\;\sigma$
& $0$&$.$&$517\, 536\;\varrho$&$\pm$&$14$&$.$&$999\, 953\;\sigma$
& & &
\\
& & &
& $4.741\, 946$&$+$&\multicolumn{3}{l}{$16\:\sigma$} 
& & & & & & &
& & &
\rule[-1.3ex]{0ex}{2ex}\\
\rule[0ex]{0ex}{2.7ex}
$(\frac{1}{16}+3$&$,$&$\frac{1}{16}+3)_{c}  $ & $m_1+m_2$ 
& $-2.890\, 111$&$\pm$&$11$&$.$&$001\, 019\;\sigma$
& $0$&$.$&$482\, 461\;\varrho$&$\pm$&$5$&$.$&$000\, 047\;\sigma$
& & &
\\
& & &
& $3.091\, 673$&$\pm$&$5$&$.$&$267\, 222\;\sigma$
& $0$&$.$&$482\, 461\;\varrho$&$\pm$&$14$&$.$&$999\, 953\;\sigma$
& & &
\\
& & &
& $3.097\, 274$&$\pm$&$14$&$.$&$605\, 702\;\sigma$
& $0$&$.$&$528\, 311\;\varrho$&$\pm$&$11$&$.$&$000\, 000^{+}\sigma$
& & &
\rule[-1.3ex]{0ex}{2ex}\\
\rule[0ex]{0ex}{2.7ex}
$(\frac{1}{2}+3$&$,$&$\frac{1}{2}+3)$         & $m_1+m_4$ 
& $-3.447\, 863$&$\pm$&$6$&$.$&$844\, 587\;\sigma$
& $0$&$.$&$485\,373\;\varrho$&$\pm$&$0$&$.$&$964\, 059\;\sigma$
& $0.003\, 800$&--&$0.003\, 850$
\\
& & & 
& $-3.399\, 457$&$+$&\multicolumn{3}{l}{$16\:\sigma$} 
& $0$&$.$&$486\, 904\;\varrho$&$\pm$&$7$&$.$&$000\, 017\;\sigma$
& & &
\\
& & & 
& $-3.048\, 514$&$\pm$&$11$&$.$&$049\, 729\;\sigma$
& $0$&$.$&$486\, 904\;\varrho$&$\pm$&$12$&$.$&$999\, 983\;\sigma$
& & &
\\
& & & 
& $3.054\, 397$&$\pm$&$11$&$.$&$080\, 356\;\sigma$
& $0$&$.$&$500\, 000^{+}\varrho$&$\pm$&$11$&$.$&$000\, 000^{+}\sigma$
& & &
\\
& & & 
& & & & & 
& $0$&$.$&$500\, 000^{+}\varrho$&$+$&\multicolumn{3}{l}{$16\:\sigma$} 
& & &
\rule[-1.3ex]{0ex}{2ex}\\
\end{tabular}
\end{table}

\widetext
\squeezetable
\begin{table}
\caption{Ratios of scaled gaps for three values 
of the scaling parameter $\mu$.}
\label{t:rat}\nopagebreak
\begin{tabular}{rrllllllllll}
\multicolumn{1}{c}{$\rule[-1.3ex]{0ex}{4ex}\mu$} &
\multicolumn{1}{c}{$N$} &
\multicolumn{1}{c}{$R_1$} &
\multicolumn{1}{c}{$R_2$} &
\multicolumn{1}{c}{$R_3$} &
\multicolumn{1}{c}{$R_4$} &
\multicolumn{1}{c}{$R_5$} &
\multicolumn{1}{c}{$R_6$} &
\multicolumn{1}{c}{$R_7$} &
\multicolumn{1}{c}{$R_{8a}$} &
\multicolumn{1}{c}{$R_{8b,c}$} &
\multicolumn{1}{c}{$R_{9}$} \\
\hline
\rule[0ex]{0ex}{2.7ex}
   &  10 & 1.618$\,$749 & 1.983$\,$943 & 2.403$\,$558 & 2.083$\,$461 
         & 2.616$\,$910 & 2.910$\,$107 & 2.317$\,$908 & 3.123$\,$561
         & 2.811$\,$391 & 3.360$\,$590 \\
   &  20 & 1.618$\,$037 & 1.983$\,$923 & 2.403$\,$649 & 2.081$\,$017 
         & 2.617$\,$335 & 2.930$\,$942 & 2.320$\,$478 & 3.086$\,$893
         & 2.781$\,$193 & 3.369$\,$241 \\
10 &  50 & 1.618$\,$024 & 1.983$\,$977 & 2.403$\,$797 & 2.079$\,$804 
         & 2.617$\,$468 & 2.934$\,$096 & 2.314$\,$580 & 3.078$\,$762
         & 2.773$\,$762 & 3.375$\,$544 \\
   &  75 & 1.618$\,$024 & 1.983$\,$984 & 2.403$\,$815 & 2.079$\,$550 
         & 2.617$\,$482 & 2.934$\,$583 & 2.313$\,$631 & 3.077$\,$950
         & 2.772$\,$773 & 3.376$\,$215 \\
\rule[-1.3ex]{0ex}{2ex}
   & 100 & 1.618$\,$025 & 1.983$\,$990 & 2.403$\,$834 & 2.079$\,$551 
         & 2.617$\,$497 & 2.934$\,$617 & 2.313$\,$523 & 3.077$\,$537
         & 2.772$\,$605 & 3.376$\,$509 \\
\hline
\rule[0ex]{0ex}{2.7ex}
   &  10 & 1.623$\,$692 & 1.989$\,$446 & 2.412$\,$441 & 2.017$\,$536 
         & 2.623$\,$691 & 2.958$\,$322 & 2.085$\,$027 & 3.006$\,$016
         & 2.671$\,$041 & 3.411$\,$712 \\
   &  20 & 1.618$\,$144 & 1.987$\,$688 & 2.405$\,$009 & 2.025$\,$316 
         & 2.618$\,$144 & 2.954$\,$218 & 2.129$\,$306 & 3.006$\,$016
         & 2.669$\,$638 & 3.404$\,$917 \\
20 &  50 & 1.618$\,$034 & 1.987$\,$597 & 2.404$\,$866 & 2.026$\,$212 
         & 2.618$\,$034 & 2.954$\,$289 & 2.132$\,$554 & 3.003$\,$342
         & 2.669$\,$034 & 3.404$\,$805 \\
   &  75 & 1.618$\,$034 & 1.987$\,$591 & 2.404$\,$865 & 2.026$\,$286 
         & 2.618$\,$034 & 2.954$\,$306 & 2.132$\,$780 & 3.003$\,$242
         & 2.668$\,$960 & 3.404$\,$808 \\
\rule[-1.3ex]{0ex}{2ex}
   & 100 & 1.618$\,$034 & 1.987$\,$589 & 2.404$\,$865 & 2.026$\,$315 
         & 2.618$\,$034 & 2.954$\,$311 & 2.132$\,$871 & 3.003$\,$212
         & 2.668$\,$941 & 3.404$\,$808 \\
\hline
\rule[0ex]{0ex}{2.7ex}
   &  10 & 1.634$\,$823 & 1.992$\,$979 & 2.428$\,$402 & 2.003$\,$678 
         & 2.634$\,$823 & 2.970$\,$263 & 2.021$\,$663 & 2.994$\,$370
         & 2.649$\,$653 & 3.428$\,$399 \\
   &  20 & 1.618$\,$424 & 1.988$\,$671 & 2.405$\,$376 & 2.012$\,$146 
         & 2.618$\,$424 & 2.956$\,$329 & 2.071$\,$838 & 2.992$\,$915
         & 2.644$\,$399 & 3.405$\,$376 \\
30 &  50 & 1.618$\,$036 & 1.988$\,$497 & 2.404$\,$869 & 2.013$\,$483 
         & 2.618$\,$036 & 2.956$\,$009 & 2.078$\,$248 & 2.992$\,$952
         & 2.645$\,$130 & 3.404$\,$869 \\
   &  75 & 1.618$\,$034 & 1.988$\,$489 & 2.404$\,$867 & 2.013$\,$604 
         & 2.618$\,$034 & 2.956$\,$007 & 2.078$\,$742 & 2.992$\,$957 
         & 2.645$\,$220 & 3.404$\,$867 \\
\rule[-1.3ex]{0ex}{2ex}
   & 100 & 1.618$\,$034 & 1.988$\,$486 & 2.404$\,$867 & 2.013$\,$644 
         & 2.618$\,$034 & 2.956$\,$007 & 2.078$\,$950 & 2.992$\,$958
         & 2.645$\,$248 & 3.404$\,$867 \\
\hline
$\rule[-1.3ex]{0ex}{4ex}\infty$ & 
$\infty$ & 1.618$\,$034 & 1.989$\,$044 & 2.404$\,$867 & 2.000$\,$000 
         & 2.618$\,$034 & 2.956$\,$295 & 2.000$\,$000 & 2.989$\,$044
         & 2.618$\,$034 & 3.404$\,$867 \\
\end{tabular}
\end{table} 

\end{document}